\documentclass[11pt]{article}
\usepackage{amsmath, amsfonts, amssymb, amsthm, mathabx}
\usepackage[T1]{fontenc}
\usepackage[utf8]{inputenc}
\usepackage[english]{babel}
\usepackage{graphicx}
\usepackage{float}
\usepackage{booktabs}
\usepackage{epsfig}
\usepackage{hyperref}
\hypersetup{
    colorlinks=true,
    linkcolor=blue,
    filecolor=magenta,      
    urlcolor=cyan,
}
\usepackage{accents}
\usepackage{comment}
\usepackage [autostyle, english = american]{csquotes}
\usepackage[percent]{overpic}

\usepackage{fancyhdr}
\usepackage{longtable}
\usepackage{tabularx}
\usepackage{ltablex}
\usepackage{subfig}

\usepackage[dvipsnames]{xcolor}
\definecolor{astral}{RGB}{0,164,239}
\usepackage{multirow} 
\usepackage{multicol} 
\usepackage[font=small,margin=0.25in,labelfont={sc},labelsep={colon}]{caption}

\usepackage{listings}

\usepackage{booktabs} 
\usepackage{siunitx}
\MakeOuterQuote{"}
\numberwithin{equation}{section}
\theoremstyle{plain}
\newtheorem*{theorem*}{Theorem}
\newtheorem*{lemma*}{Lemma}
\newtheorem{theorem}{Theorem}
\newtheorem{lemma}{Lemma}[section]

\theoremstyle{definition}

\usepackage{geometry}
\begin{document}

\title{Parameter Estimation and Adaptive Solution of the Leray-Burgers Equation using Physics-Informed Neural Networks}

\author{
Bong-Sik Kim~\footnote{Corresponsing author: Department of Mathematics and Physics, 
American university of Ras Al Khaimah, UAE. 
E-mail: bkim@aurak.ac.ae},\ 
\ 
Yuncherl Choi~\footnote{Ingenium College of Liberal Arts,
Kwangwoon   University,  Seoul  01891, Korea.
E-mail : yuncherl@kw.ac.kr},\ 
and \ 
Doo Seok Lee~\footnote{Department of Undergraduate Studies,
Daegu Gyeongbuk Institute of Science and Technology,
Daegu  42988, Korea.
E-mail : dslee@dgist.ac.kr}
}

\date{\today}

\maketitle

\begin{abstract}
This study presents a unified framework that integrates physics-informed neural networks (PINNs)
to address both the inverse and forward problems of the one-dimensional Leray-Burgers equation. 
First, we investigate the inverse problem by empirically determining the characteristic
wavelength parameter $\alpha$ at which the Leray-Burgers solutions closely approximate
those of the inviscid Burgers equation. Through PINN-based computational experiments on 
inviscid Burgers data, we identify a physically consistent range for $\alpha$ being between 0.01 and 0.05 for continuous initial conditions and between 0.01 and 0.03 for discontinuous profiles, demonstrating the dependence of $\alpha$ on initial data.
Next, we solve the forward problem using a PINN architecture where $\alpha$ is 
dynamically optimized during training via a dedicated subnetwork, Alpha2Net. Crucially, 
Alpha2Net enforces $\alpha$ to remain within the inverse problem-derived bounds, ensuring 
physical fidelity while jointly optimizing network parameters (weights and biases). 
This integrated approach effectively captures complex dynamics, such as shock and 
rarefaction waves.
This study also highlights the effectiveness and efficiency of the Leray-Burgers
equation in real practical problems, specifically Traffic State Estimation.
\end{abstract}

\section{Introduction}\label{sec:introduction}

In this study, we explore the one-dimensional Leray-Burgers (LB) equation \eqref{eq-main-a}, 
a regularized model of the inviscid Burgers equation. 
The equation introduces a wavelength parameter \(\alpha\)
to prevent finite-time blow-ups while aiming to preserve the essential dynamics of 
the inviscid case, such as shock waves.
However, selecting an appropriate $\alpha$ is non-trivial;
it significantly impacts the solution's fidelity to the physical, 
inviscid dynamics. An incorrect $\alpha$ leads to over-smoothing or insufficient regularization. 
This sets up the need for a systematic approach.
To address this challenge, 
we employ Physics-Informed Neural Networks (PINNs) \cite{Raissi1, Raissi2, GYL, DD1, DD2}
in a two-step process that bridges the inverse and forward problems.
 
First, we tackle the inverse problem:
we leverage PINNs to systematically and empirically determine a physically 
consistent range for the parameter $\alpha$. This range ensures that the LB solution accurately 
approximates the entropy solution of the inviscid Burgers equation under various initial conditions,
establishing a foundation of physical fidelity
 (Section \ref{Sec:Inverse}). 
We find that the choice of \(\alpha\) depends on the initial data.
For continuous initial profiles, the practical range of \(\alpha\) is \(0.01\)--\(0.05\), whereas for discontinuous initial profiles, it is \(0.01\)--\(0.03\) 
(\ref{section-inviscid}).
This step establishes a critical foundation by identifying the bounds within which $\alpha$ yields physically meaningful results.

Second, we address the forward problem with a unique adaptive architecture
  in Section \ref{section-data-driven}. 
We introduce Alpha2Net, a dedicated subnetwork designed to dynamically learn the optimal 
time-dependent parameter, $\alpha(t)$, during training. 
A key innovation of our approach is that Alpha2Net is explicitly constrained to operate 
within the physically meaningful bounds established by our inverse problem study.
This synergistic design ensures that the dynamic adaptation remains physically grounded, 
preventing the network from converging to non-physical or unstable solutions.

To demonstrate the practical relevance and efficiency of our framework, we apply the Leray-type 
regularization to a real-world problem in Section \ref{section-TSE}: 
Traffic State Estimation (TSE). 
We introduce an LWR-$\alpha$ model, a variant of the Lighthill-Whitham-Richards (LWR) traffic model 
based on the Leray-Burgers equation.
Our results show that the LWR-$\alpha$ model, using an $\alpha$ value consistent with our inverse 
problem findings, not only captures the complex, nonlinear dynamics of traffic flow but also proves
to be significantly more computationally efficient than standard or viscous-based models. This application serves as a strong validation of our integrated framework.

\section{Background}\label{sec:background}

\subsection{Leray-Burgers Equation}

 We consider a problem of computing the solution $v: [0,T] \times \Omega \to \mathbb{R}$ 
 of an evolution equation
\begin{align}
     v_t(t,x) + \mathcal{N}_\alpha\,[v](t,x) = 0, 
       &\quad  \forall (t,x) \in [0,T] \times \Omega, \label{eq-main-a}\\
    v(0,x) = v_0(x), \quad &\quad \forall x \in \Omega, \nonumber
\end{align}
where $\mathcal{N}_\alpha$ is a nonlinear differential operator acting on $v$ 
with a small constant parameter $\alpha >0$, 
 \begin{equation}\label{nonlinear-term}
   \mathcal{N}_\alpha[v] = vv_x + \alpha^2v_xv_{xx}.
 \end{equation}
Here, 
$\Omega \subset \mathbb{R}$ is a bounded domain,
$T$ denotes the final time and
$v_0: \Omega \to \mathbb{R}$ is the prescribed initial data.
Although the methodology allows for different types of boundary conditions, 
we restrict our discussion to  Dirichlet or periodic cases
 and prescribe the boundary data as
\[
       v_b(t,x) = v(t,x), \quad \forall (t,x) \in [0,T] \times \partial\Omega. 
\]       

Equation \eqref{eq-main-a} is called the Leray-Burgers equation (LB). 
It is also known as
   \emph{Burgers-$\alpha$}, \emph{connectively filtered Burgers equation}, 
   \emph{Leray regularized reduced order model}, etc.,
   in literature.
 Bhat and Fetecau \cite{BF1} introduced \eqref{eq-main-a} as 
 a regularized approximation to 
 the inviscid Burgers equation
\begin{equation}\label{BE}
   v_t + vv_x = 0. 
\end{equation}       
They considered a special smoothing kernel associated 
with the Green function of the Helmholtz operator
\[ u_\alpha = \mathcal{H}^{-1}_\alpha v = (I - \alpha^2\partial_x^2)^{-1}v, \ \ (I = identity),\]
where $\alpha$ 
is interpreted as the \emph{characteristic wavelength scale} 
            below which the smaller physical phenomena
            are averaged out
            and it accelerates energy decay  \cite{Holm2, Norgard}.
Applying the smoothing kernel to the convective term in \eqref{BE}
yields
\begin{equation}\label{B_alpha}
   v_t + u_\alpha v_x = 0,
\end{equation}
where $v = v(t,x)$ is a \emph{vector field} and  $u_\alpha$ is the \emph{filtered vector field}.
The filtered vector $u_\alpha$ is smoother than $v$ and
the equation (\ref{B_alpha}) is a nonlinear Leray-type regularization \cite{Leray} 
of the inviscid Burgers equation.   
Here and in the following, we abuse the notation of the filtered vector $u_\alpha$ with $u$.
If we express the equation (\ref{B_alpha}) in the filtered vector $u$, 
it becomes a quasilinear evolution equation that consists of the inviscid Burgers equation plus 
$\mathcal{O}(\alpha^2)$ nonlinear terms \cite{BF1, BF2, BF3}:
\begin{equation}\label{eq1.old}
 u_t + u u_x =   \alpha^2 (u_{txx} +  u u_{xxx}).
\end{equation}
In this paper,  we follow Zhao and Mohseni \cite{ZM} to expand the
inverse Helmholz operator in \(\alpha\) to higher orders of the
Laplacian operator: 
 \[
  (1-\alpha^2\Delta )^{-1} = 1+\alpha^2\Delta + \alpha^4\Delta^2 + \cdots \ 
     \mbox{if}\ \alpha\lambda_{\mathrm{max}} < 1,
\] 
where \(\lambda_{\mathrm{max}}\) is the highest eigenvalue of the
discretized operator \(\Delta\). Then we can write \eqref{B_alpha} in the unfiltered
vector fields $v$ to obtain the equation \eqref{eq-main-a}-\eqref{nonlinear-term} with $\mathcal{O}(\alpha^4)$
truncation error.

The inviscid Burgers equation develops shocks--discontinuities 
where solution gradients become infinite--in finite time,
even from smooth initial data. 
 This occurs due to the intersection of characteristic curves.
 The regularization introduced by $\alpha$ counteracts this by introducing nonlinear, 
 dispersive-like effects via \eqref{nonlinear-term}, which become significant 
 where gradients $v_x$ are large, near incipient shocks. 
 This regularization smooths sharp fronts, preventing gradient blow-up 
 by bending characteristics to avoid intersection \cite{BF1, BF3}, 
 ensuring globally existing smooth solutions for any $\alpha > 0$,
as established by Bhat and Fetecau in  Section 2 of \cite{BF1}:
  
\begin{theorem}
  Given initial data $v_0 = v(0,x)\,\in\,W^{2,1}(\mathbb{R})= 
    \{ u\in L^1(\mathbb{R}): D^su\in L^1(\mathbb{R}) \ \ \mathit{for\ all}\ \  |s|\leq 2\}$, 
    there exists a unique solution
    $v(t,x)\in W^{2,1}(\mathbb{R})$ for all $t>0$ to the Leray-Burgers equation \eqref{B_alpha}. 
\end{theorem}

\noindent 
Furthermore, the Leray-Burgers solution $u_{\alpha}(t,x)$  with initial data
$
 u_{\alpha}(0,x) = \mathcal{H}^{-1}_{\alpha}v_0(x)
 $ for
$ v_0 \in W^{2,1}(\mathbb{R}) 
$
converges strongly, as $\alpha\rightarrow 0^+$, to  a global weak solution $v(t,x)$
of the following initial value problem for the inviscid Burgers equation (Theorem 2 in \cite{BF1}):
\begin{align*}   
v_t + \frac{1}{2}\left(v^2\right)_x = 0 \ \ \  \mbox{with}\ \ 
      v(0,x) = v_0(x). 
\end{align*}      
Bhat and Fetecau \cite{BF1} found numerical evidence that the chosen weak solution in the zero-$\alpha$ limit satisfies
the Oleinik entropy inequality, making the solution physically appropriate.
Thus, the $\alpha$-regularization not only tames the mathematical difficulties 
of shocks but also guides the solution towards physical fidelity in the inviscid limit.
The proof relies on uniform estimates of the unfiltered velocity $v$ rather than the filtered velocity $u$. It made 
possible the strong convergence of the Leray-Burgers solution to the correct entropy solution
of the inviscid Burgers equation. 
In the context of the filtered velocity $u_\alpha$,
 they also showed that the Leray-Burgers equation captures the correct shock solution of the inviscid Burgers equation 
 for Riemann data consisting of a single decreasing jump  \cite{BF3}. 
 However, since  $u_\alpha$ captures an unphysical solution for Riemann data comprised of a single increasing jump, 
 it was necessary to control the behavior of the regularized equation by introducing an arbitrary mollification of the Riemann data 
 to capture the correct rarefaction solution of the inviscid Burgers equation. 
 With that modification, they extended the existence results to the case of discontinuous initial data $u_\alpha \in L^\infty$.
  However, it is still an open problem for the initial data $v_0\in L^\infty$.
  In  \cite{Guel},  Guelmame et al. derived a similar regularized equation to \eqref{eq1.old}:
  \begin{equation}\label{eq1.similar}
 u_t + u u_x =   \alpha^2 (u_{txx} + 2u_x u_{xx} + u u_{xxx}),
\end{equation}
  which has an additional term $ 2u_xu_{xx}$ on the right-handed side. 
  Notice that $u$ in this equation 
  is the filtered vector field in \eqref{B_alpha}. 
  When they were establishing the existence of the entropy solution, Guelmame et al.
  resorted to altering the equation 
  \eqref{eq1.similar}, as Bhat and Fetecau had to modify 
  the initial data for their proof in \cite{BF3}.
  Analysis in the context of the filtered vector field $u$ appears 
  to induce an additional modification of 
  the equations to achieve the desired results. 
  Working with the actual vector field $v$ may avoid such arbitrary changes. 
  
 Equation \eqref{B_alpha} and related models have previously appeared in the literature.
 We refer \cite{Araujo, BF1, BF2, BF3, Gottwald, Guel, ILX, Pav, SJ, Zhang, Villavert}
 for more properties related to  the Leray-Burgers  equation.
  The paper \cite{SJ} explores the role of 
 $\alpha$ in regularizing Proper Orthogonal Decomposition (POD)-Galerkin models for 
 the Kuramoto-Sivashinsky (KS) equation. The $\alpha$-regularization is 
 introduced to enhance the stability and accuracy of these models by applying Helmholtz filtering 
 to the eigenmodes of the quadratic terms. 
  The link between regularization procedures such as Helmholtz regularization and numerical
              schemes, for example, had been studied in \cite{Gottwald, Pav}. 
        They argued that, in numerical computations, 
            the parameter $\alpha^2$ cannot be interpreted solely as a length scale
            because it also depends on
             the numerical discretization scheme chosen. 
             They observed that 
             the choice of $\alpha$  depends on a relation between
             $\alpha$ and the mesh size that preserves
          stability and consistency with conservation conditions for the chosen numerical scheme
          \cite{BF1, Gottwald, Pav}.
          Also, they found that, for a fixed number of grid points,
          there is a particular value of $\alpha\approx 0.02$ below which the solution
             becomes oscillatory (even with continuous initial profiles).

\subsection{Leray Regularization and Conserved Quantities}

The Leray-type regularization, originally introduced by Jean Leray for the Navier-Stokes equations
governing incompressible fluid flow \cite{Leray}, enhances stability and accuracy 
by applying a Helmholtz filter to the convective term. 
In the Fourier domain, this corresponds to
\[ \widehat{\mathcal{H}^{-1}v }= \frac{\widehat{v}}{1+\alpha^2k^2},\]
which attenuates high-wavenumber modes typically responsible for instability.
The parameter $\alpha$ defines the subgrid length scale over which unresolved motions are averaged, 
preserving the dominant flow dynamics \cite{Holm2}. 
This spatial averaging acts as a regularization mechanism, consistent with Leray’s original 
mollification approach \cite{Leray} to ensure well-posedness. 
The resulting suppression of small-scale features can also lead to accelerated energy decay, 
making the system more dissipative than the underlying inviscid model \cite{Norgard}.

In the realm of stochastic partial differential equations, Leray regularization has 
proven effective in enhancing the stability of reduced order models (ROMs), 
particularly for convection-dominated systems. Iliescu et al. \cite{ILX}
explored this in their study of a stochastic Burgers equation driven by linear multiplicative 
noise. They found that standard Galerkin ROMs (G-ROMs) produce spurious numerical oscillations 
in convection-dominated regimes, a problem exacerbated by increasing noise amplitude. 
To counter this, they applied an explicit spatial filter to the convective term creating 
the Leray ROM (L-ROM). This approach significantly mitigates oscillations, yielding more 
accurate and stable solutions compared to the G-ROM, especially under stochastic perturbations.
The L-ROM’s robustness to noise variations suggests that Leray regularization may help preserve
statistical properties or conserved quantities, such as energy or moments of the solution, in a stochastic context. This extends the utility of Leray regularization beyond deterministic settings, offering a practical tool for modeling complex stochastic dynamics while maintaining numerical fidelity.

Beyond its stabilizing effects in both deterministic and stochastic frameworks, 
the Leray regularization, as embodied in the Leray-Burgers equation \eqref{eq-main-a} 
(or \eqref{eq:Leray-Burgers}), 
also impacts the conservation of key physical quantities. The subsequent lemmas establish 
the conditions under which energy and mass are conserved, hinging critically on the spatial 
independence of the regularization parameter $\alpha(t)$.

\begin{lemma}[Conservation of Energy]
Let \( v(t,x) \) satisfy the Leray-Burgers equation
\begin{equation} \label{eq:Leray-Burgers}
    v_t + v\,v_x + \alpha^2 v_x v_{xx} = 0,
\end{equation}
on a periodic domain \( a \le x \le b \) and for time \( t \ge 0 \).
If \( \alpha = \alpha(t) \) is independent of \( x \), then the total energy
\[
E(t) := \int_a^b \left( \tfrac{1}{2} v^2 + \tfrac{1}{2} \alpha^2 v_x^2 \right) dx
\]
is conserved in time. \label{lemma2.1}
\end{lemma}

\begin{proof}
We first rewrite equation \eqref{eq:Leray-Burgers} in the form
\[
v_t + \frac{\partial}{\partial x} \left( \tfrac{1}{2} v^2 + \tfrac{1}{2} \alpha^2 v_x^2 \right) - \tfrac{1}{2} (\alpha^2)_x v_x^2 = 0.
\]
If \( \alpha = \alpha(t) \), then \( (\alpha^2)_x = 0 \), and the equation becomes
\[
v_t + \frac{\partial}{\partial x} \left( \tfrac{1}{2} v^2 + \tfrac{1}{2} \alpha^2 v_x^2 \right) = 0.
\]
Integrating over \( [a, b] \) and using periodic boundary conditions, we obtain
\[
\int_a^b \left( \tfrac{1}{2} v^2 + \tfrac{1}{2} \alpha^2 v_x^2 \right) dx = C
\]
for an arbitrary constant $C$
so that 
\[ \frac{dE}{dt} = 0, \]
which confirms the conservation of energy.
\end{proof}

\begin{lemma}[Conservation of Mass]
Under the same assumptions as in Lemma \ref{lemma2.1}, the total mass
\[
M(t) = \int_a^b v(t,x) \, dx
\]
is conserved in time.
\end{lemma}

\begin{proof}
Integrating equation \eqref{eq:Leray-Burgers} over \( [a, b] \), we find
\[
\frac{d}{dt} \int_a^b v(t,x) \, dx = \int_a^b v_t \, dx = - \int_a^b \frac{\partial}{\partial x} \left( \tfrac{1}{2} v^2 + \tfrac{1}{2} \alpha^2 v_x^2 \right) dx.
\]
Since the integrand is a spatial derivative and the boundary conditions are periodic, the integral vanishes:
\[
\frac{dM}{dt} = 0.
\]
Hence, mass is conserved.
\end{proof}

Both conservation properties rely critically on the assumption that \( \alpha = \alpha(t) \), i.e., the regularization parameter is spatially uniform. If \( \alpha \) depends on \( x \), the equation contains a non-vanishing source term that may lead to dissipation or amplification of energy and may violate mass conservation as well.

\section{PINN Structure for Inverse and Forward Problems}

\begin{figure}[h]
\begin{center}
\includegraphics[width=0.7\textwidth]{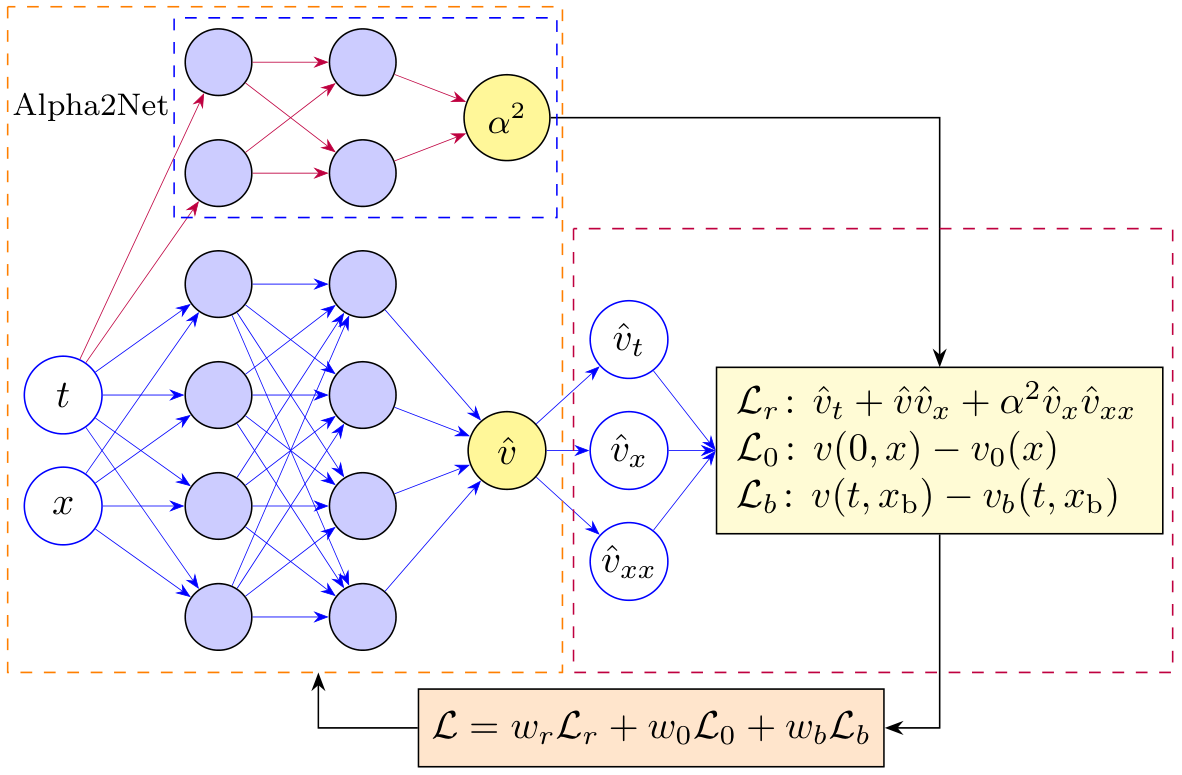}
\end{center}
\caption[Fig1]{\small The PINN architecture for solving the 
LB equation: the diagram illustrates the surrogate neural network for predicting 
$v(t,x)$, the Alpha2Net subnetwork for optimizing 
$\alpha$, and the loss components enforcing the PDE, initial, and boundary conditions.}
\label{fig1}
\end{figure}

We employ a Physics-Informed Neural Network (PINN) to address 
both the inverse and forward problems for the Leray-Burgers (LB) equation, 
as depicted in Figure \ref{fig1}. The LB equation, given by
\begin{align*}
  v_t+vv_x + \alpha^2 v_x v_{xx} = 0, 
\end{align*}
is solved in various initial and boundary condition scenarios, with the characteristic wavelength parameter $\alpha$,
fixed ($\alpha = \text{constant}$) or 
adaptively optimized ($\alpha = \alpha(t)$).
The PINN architecture, shown in Figure \ref{fig1}, consists of two primary 
components: a main neural network to approximate the solution 
$v(t,x)$  and a subnetwork, Alpha2Net, to learn the parameter 
$\alpha$.
The main network is a fully connected feed-forward neural network 
(multilayer perceptron, MLP) with eight hidden layers, each containing 20 neurons, 
using $\tanh$ activation functions.
Following standard practice in the PINN literature—including seminal works 
by Raissi et al. \cite{Raissi1, Raissi2} on the Burgers equation—we adopt this architecture 
for its effective balance between expressiveness and trainability in PDE problems.
To ensure its suitability for the Leray-Burgers equation, we conducted preliminary tests with 
alternative configurations (e.g., varying depth and width). 
The chosen [8 layers, 20 neurons] setup showed reliable convergence and accuracy in minimizing 
the $L^2$ error and PDE residual during the inverse problem (Section 4), 
without incurring excessive computational cost or overfitting. 
For consistency, this architecture—trained for 10,000 epochs the inverse problem 
and 20,000 epochs in the forward inference problem —was used across all models presented,
offering a robust and comparable foundation even if not individually optimized.
The network takes spatio-temporal inputs $(t,x)$ 
 and outputs the predicted solution $v(t,x)$.
To enforce the physics of the LB equation, automatic differentiation is used to compute 
the required derivatives $\hat{v}_t$, $\hat{v}_x$, and $\hat{v}_{xx}$,
which are then used to evaluate the PDE residual.

The Alpha2Net subnetwork, highlighted in the upper part of Figure \ref{fig1}, 
is designed to adaptively learn the parameter  $\alpha$ as a function of time, 
i.e. $\alpha(t)$. 
This choice, while simplifying the learning task for Alpha2Net, 
is motivated by several factors. Firstly, a spatially uniform $\alpha(t)$ ensures that 
the Leray-Burgers equation retains important conservation properties for mass and 
a defined energy (as demonstrated in Lemmas 2.1 and 2.2), which can be compromised 
if $\alpha$ varies spatially. Secondly, it allows $\alpha(t)$ to act as a global, 
time-evolving regularization scale, adapting to the overall characteristics of the solution 
as it develops. While future work might explore spatially varying or gradient-dependent $\alpha$
for potentially more localized adaptation, the current $\alpha(t)$ formulation provides 
a robust and interpretable framework for dynamic regularization.
This subnetwork is a smaller MLP with three hidden layers, 
each containing 10 neurons, and also uses $\tanh$ activation functions. 
This relatively compact architecture was chosen for Alpha2Net 
as it learns a scalar function $\alpha(t)$ dependent only on time, 
a task generally not requiring the extensive capacity of the main solution network. 
Such an architecture provides sufficient expressiveness for this auxiliary task 
while minimizing additional computational overhead and was observed to effectively learn stable 
and physically-informed $\alpha(t)$ profiles during training.
It takes the time coordinate $t$ as its sole input and outputs $\alpha^2(t)$,
 for use in the PDE residual. 
 To ensure physical consistency, Alpha2Net constrains $\alpha^2(t)$ to lie within the range $[10^{-4}, 0.01]$, 
 which is slightly broader than the practical range identified in the inverse problem (Section~\ref{Sec:Inverse}). 
 This modest extension grants Alpha2Net additional flexibility 
 during the dynamic learning process, allowing it to explore values that may temporarily 
 aid in stabilizing training or navigating complex solution features, 
 while remaining anchored by a physically-informed lower bound and a reasonable upper limit 
 that prevents excessive smoothing.

  Although a rigorous error analysis for PINNs with adaptive parameters 
  such as $\alpha(t)$ remains an open research question, the imposed constraints are 
  designed to guide the optimization toward a regime where error is expected to be controlled. 
  The lower bound enhances stability and promotes the well-posedness of the learning problem, 
  while the upper bound ensures the solution does not diverge significantly 
  from the target physics due to over-regularization. This constrained range has also been 
  empirically validated in the inverse problem, yielding solutions with low discrepancy 
  from known references.

The constraint is implemented using a sigmoid activation function at the output layer of Alpha2Net, scaled to map the output $z$ to the desired range:
 $$ \alpha^2(t) = 10^{-4} + (0.01-10^{-4})\cdot \text{sigmoid}(z),
 $$
where $z$ is the raw output of the subnetwork. This ensures that $\alpha(t)$
remains within the specified bounds during training, preventing the network 
from converging to nonphysical values.

The output of the main network $(\hat{v}, \hat{v}_t, \hat{v}_x, \hat{v}_{xx})$
and Alpha2Net($\alpha$) are combined to compute the PDE residual, 
as shown in the right part of Figure \ref{fig1}.
The loss function is designed to enforce the physics of the LB equation and 
consists of three key components:
\begin{itemize}
    \item \textbf{Residual Loss} (enforcing the PDE):
    \[
    \mathcal{L}_{\text{r}} = \frac{1}{N_r} \sum_{i=1}^{N_r} \left( \frac{\partial \hat{v}}{\partial t} + \hat{v} \frac{\partial \hat{v}}{\partial x} + 
     \alpha^2 \frac{\partial^2 \hat{v}}{\partial x} \frac{\partial^2 \hat{v}}{\partial x^2} \right)^2,
    \]
    where $N_r$
  is the number of collocation points sampled across the spatiotemporal domain.
    \item \textbf{Initial Condition Loss}:
    \[
    \mathcal{L}_{\text{0}} = \frac{1}{N_0} \sum_{i=1}^{N_0} \left( \hat{v}(0, x_i) - v_0(x_i) \right)^2,
    \]
    where $N_0$
  is the number of points sampled along the initial condition at $t=0$.
    \item \textbf{Boundary Condition Loss}:
    \[
    \mathcal{L}_{\text{b}} = \frac{1}{N_b} \sum_{i=1}^{N_b} 
       \left( \hat{v}(t_i, x_{\text{b}}) - v_b(t_i, x_{\text{b}}) \right)^2,
    \]
    where $N_b$
  is the number of points sampled along the boundary $x=x_b$.
\end{itemize}
These terms are combined into a total loss 
\[ \mathcal{L} = 
w_r \mathcal{L}_{\text{r}} + w_0 \mathcal{L}_{\text{0}} 
+ w_b \mathcal{L}_{\text{b}},
\]
where the weights \( w_r, w_0, w_b \) are either fixed (e.g., set to 1 for equal weighting)
or tuned dynamically during training to balance the contributions of each loss term.
The training points are adaptively sampled using Latin Hypercube Sampling, with a focus on regions exhibiting high PDE residuals or steep solution gradients, such as shock regions near discontinuities in the initial conditions.

The model is optimized using either the ADAM or Limited-Memory BFGS (L-BFGS) optimizer
with a decaying learning rate schedule over unit epochs. 
During training, the parameters of both the main network 
(weights and biases) and Alpha2Net are updated simultaneously to minimize the total loss
$\mathcal{L}$.
Performance is assessed by computing 
the $L^2$-error between the PINN predictions and the analytical solutions of
the inviscid Burgers equation, obtained via the method of characteristics.

The concept of adaptively learning regularization or artificial viscosity parameters within PINNs 
has been explored to tackle challenges with hyperbolic PDEs. 
For instance, Coutinho et al. \cite{Coutinho} proposed methods to learn a global or 
localized artificial viscosity coefficient $\nu$
for a standard diffusion, $\nu v_{xx}$, added to stabilize equations like 
the inviscid Burgers equation.
Our approach, Alpha2Net, takes a fundamentally different route and 
offers specific advantages  for the Leray-Burgers equation. 
Rather than adding a generic linear artificial viscosity, 
 Alpha2Net learns the time-dependent  parameter $\alpha(t)$ 
 governing the intrinsic nonlinear regularization term 
 $\alpha^2 v_x v_{xx}$,  derived from the Leray filtering framework.  
 This term has distinct mathematical properties and physical meaning, 
as it represents a characteristic length scale $\alpha$, 
in contrast to artificial viscosity.
Thus, we adapt a parameter within a theoretically derived, physically grounded regularized model, 
rather than injecting an external stabilization term that may distort solution behavior. 
As shown in Figure 1, Alpha2Net integrates physical constraints directly into 
the neural architecture, enabling the simultaneous approximation of $v(t, x)$ 
and adaptive tuning of $\alpha(t)$. 
This ensures physically meaningful solutions and distinguishes our method 
from those that merely learn artificial viscosity coefficients.

\section{Inverse Problem for the Estimation of Parameter \texorpdfstring{$\alpha$}{alpha}} \label{Sec:Inverse}

We  set up the computational  frame for the governing system \eqref{eq-main-a}  by 
\begin{eqnarray} 
   v_t + \lambda_1vv_x + \lambda_2v_xv_{xx} = 0,\ \ t\in [0, T],\ x\in \Omega&  \label{PDE-a} & \\
   v(0,x) = f(x),\ x\in \Omega \hspace{.5in} \label{PDE-b}  & & \\
   v(t,x) = g(t,x),\ t\in [0, T],\ x\in \partial\Omega, \hspace{0.1in} \label{PDE-c}  & &
\end{eqnarray}
where $\Omega\subset \mathbb{R}$ is a bounded domain, $\partial\Omega$ is 
a boundary of $\Omega$, $f(x)$ is an initial distribution,
and $g(t,x)$ is a boundary data. 
We intentionally introduced a new parameter $\lambda_1$ and  
set $\lambda_2 = \alpha^2$. 
During the training process, the PINN will learn $\lambda_1$ to determine the validity of the obtained $\alpha$
for the inviscid Burgers equation  ($\lambda_1 = 1$) along with the relative errors.  
We use numerical or analytical  solutions
of the exact inviscid and viscous Burgers equations  to
generate training data sets $D$ with different initial and boundary conditions:
  \[ D = \left\{ (t_i, x_i, v_i), \ i = 1, ..., N_d \right\},
   \]
  where $v_i = v(t_i,x_i)$ denotes the output value at position $x_i\in\Omega$
  and time $0<t_i\leq T$ with the final time $T$.
  $N_d$ refers to the number of training data.
  Our goal is to estimate the effective range of $\alpha$ such that
  the neural network  $v_\theta$ satisfies the equation \eqref{PDE-a}-\eqref{PDE-c}
  and $v_\theta (t_i, x_i) \approx v_{i}$. 
 The selected training models  represent a  range of initial conditions, from continuous initial data to discontinuous data,
 displaying shock and rarefaction waves.

\subsection{PINN for Inverse Problem }

Following the original work of Raissi et al.  \cite{Raissi1, Raissi2},
we use a Physics-Informed Neural Network (PINN) to determine physically meaningful 
$\alpha$-values closely approximating the entropy solutions to the inviscid Burgers equation.
 For the inverse problem, Alpha2net in Figure \ref{fig1}
 is not used because we are looking for a fixed value of $\alpha$. 
 The PINN enforces the physical constraint, 
  $$
  \mathcal{F}(t,x) := v_t + \lambda_1vv_x + \lambda_2v_xv_{xx} 
  $$ 
   on the MLP surrogate $\hat{v}(t,x) = v_\theta(t,x;\xi)$,
  where   $\theta = \theta (W,b)$ denotes all parameters of the network 
  (weights $W$ and biases $b$) and $\xi = (\lambda_1, \lambda_2)$ the physical parameters
  in \eqref{PDE-a}, acting directly in the loss function
 \begin{equation}\label{pinn-loss}
   \mathcal{L}(\theta,\xi) = \mathcal{L}_d(\theta, \xi) + \mathcal{L}_r(\theta, \xi), 
 \end{equation}
 where $\mathcal{L}_d$ is the loss function on the available measurement data set
 that consists in the mean-squared-error (MSE) between
 the MLP's predictions and training data and $\mathcal{L}_r$ is the additional residual
 term quantifying the discrepancy of the neural network surrogate $v_\theta$ with
 respect to the underlying differential operator 
 in \eqref{PDE-a}. 
 Note that $\mathcal{L}_d = w_0\mathcal{L}_0+w_b\mathcal{L}_b$ with $w_0=w_1=1$
 in Figure \ref{fig1}.
 We define 
 the {\em data residual} at $(t_i,x_i,v_i)$ in $D$:
\[
  \mathcal{R}_{d,\theta}(t_i,x_i;\xi) := v_\theta (t_i,x_i;\xi) - v_i, 
\]  
and
 the {\em PDE residual} at $(t_i,x_i)$ in $D$:
\[  \mathcal{R}_{r,\theta}(t_i,x_i;\xi) := \partial_tv_\theta
      + \lambda_1 v_\theta\partial_x v_\theta + \lambda_2\partial_x v_\theta \partial_{xx}v_\theta ,
\] 
 where $v_\theta = v_\theta(t,x;\xi)$.
Then, the data loss and residual loss functions in \eqref{pinn-loss} can be written as
\begin{align*}
  \mathcal{L}_d(\theta,\xi) 
       &= \frac{1}{N_d}\sum_{i=1}^{N_d} \Big|\mathcal{R}_{d,\theta}(t_{i},x_{i};\xi)\Big|^2 ,\\
   \mathcal{L}_r(\theta,\xi) &=  
     \frac{1}{N_r}\sum_{i=1}^{N_r} \Big|\mathcal{R}_{r,\theta}(t_{i},x_{i};\xi) \Big|^2.
\end{align*}     
 The goal is to find the network and physical parameters $\theta$ and $\xi$ that minimize the loss function \eqref{pinn-loss}:
\[
(\theta^*, \xi^*) = \underset{\theta\in \Theta,  \xi\in \Xi}{\arg\min}\,\mathcal{L}(\theta, \xi)
\]
over an admissible set $\Theta$ and $\Xi$ of training network parameters 
  $\theta$ and $\xi$, respectively.

In practice, given the set of scattered data $v_{i} = v(t_i, x_i)$, 
the MLP takes the coordinate \((t_i,x_i)\)
as input and produces output vectors \(v_\theta (t_i,x_i;\xi)\) that have the same dimension as \(v_i\).
The PDE residual $\mathcal{R}_{r,\theta}(t,x;\xi)$ forces
the output vector \(v_\theta\) to comply with the physics imposed by
the LB equation.
The PDE residual network takes 
its derivatives with respect to the input variables $t$ and $x$ 
by applying the chain rule to differentiate the compositions
of functions using the automatic differentiation integrated into TensorFlow. 
The residual of the underlying differential equation is evaluated using these gradients.
The data loss and the residual loss are trained 
using input from across the entire domain of interest. 

\subsection{Experiment 1: Inviscid with Riemann Initial Data}\label{section-inviscid}

We consider the inviscid Burgers equation \eqref{BE} with
some standard Riemann initial data of the form
%
\[
v_0(x) = \left\{ \begin{array}{cc}
                        v_L, & x\leq 0 \\
                      v_M, & 0 < x \leq 1\\
                      v_R, & x> 1
                  \end{array}
            \right.
\]

 We used the conservative upwind difference scheme to generate training data. 
For each initial profile, we computed   $256\times 101 = 25856$ data points throughout
the entire spatiotemporal domain.
 We modified the code  in  \cite{Raissi2} and, for each case,
 performed ten computational simulations with 2000 training data randomly sampled for each computation.
  The PINN model (as described in Section 3) was used for the experiments
  with $10000$ epochs.
 We adopted
    the Limited-Memory BFGS (L-BFGS) optimizer with a learning rate of 0.01 to minimize 
    MSE \eqref{pinn-loss}.
  When the  L-BFGS optimizer diverged, we preprocessed
  with the ADAM optimizer and finalized the optimization with the L-BFGS. 
  It is useful to use such a combined optimizer \cite{DD1,DB,DD3}.
  One remark is that 
 our problem is identifying the model parameter $\alpha$ rather than inferencing solutions, 
    and it is unnecessary to consider physical causality 
 in our loss function \eqref{pinn-loss} as pointed out in  \cite{WPS}.

Upon training, the network is calibrated to predict the entire
  solution \(v(t,x)\), as well as the unknown parameters $\theta$ and $\xi$.
    Along with the relative $L^2$-norm of the difference between the exact solution and 
  the corresponding trial solution
\[
     E_r (= E_r(\hat{v})):= \frac{||v-\hat{v}||_2}{||v||_2}, 
\]
  we used the absolute error of $\lambda_1$, $$\epsilon(\lambda_1) = |1 - \lambda_1|$$ 
  in determining the validity of each computational result.
   When appropriate, we will also measure the averaged relative $L^2$ error in time,
  \[ \bar{E}_r = \frac{1}{T} \int_0^T\frac{||v-\hat{v}||_2}{||v||_2}\,dt. \]
  The practical range of $\alpha$ was determined by ensuring the relative error
  $E_r$ remained below $10^{-2}$ while $\epsilon(\lambda_1)<0.01$,
  aligning the LB solution with the inviscid Burgers entropy solution.
  The results
show that the $\alpha$ value depends on the initial data,
with the effective range of $\alpha$ being between 0.01 and 0.05 for continuous initial profiles
and between 0.01 and 0.03 for discontinuous initial profiles.    

This empirically determined range is physically meaningful, as it reflects a balance
between fidelity and regularization. Within this range, $\alpha$ is small enough for the LB solution 
to accurately capture key features of the inviscid Burgers dynamics—such as shock formation 
and rarefaction waves—evidenced by low relative errors. At the same time, $\alpha$ remains large 
enough to suppress spurious oscillations or instabilities, which are especially problematic 
in PINN-based approximations when regularization is insufficient or when attempting to resolve 
features below the network’s effective resolution. 
Values of $\alpha$ outside this range tend to compromise the solution: overly large values 
lead to excessive smoothing and loss of physical detail, 
while overly small values result in inadequate regularization and potential numerical breakdown.

The following experiments illustrate these findings in detail.

 \subsubsection{Shock Waves}

{\small 
\begin{table}[h]
\centering
\begin{tabular}{|wc{1.6em}|wc{2.7em}|wc{2.7em}|wc{2.7em}||wc{2.7em}|wc{2.7em}|wc{2.7em}|} 
 \toprule 
   \multirow{2}{*}{No.} &  \multicolumn{3}{c||}{Initial Profile (I) } 
     &  \multicolumn{3}{c|}{Initial Profile (II)}    \\ \cmidrule{2-7}  
       & $\lambda_2\!=\!\alpha^2$ & $\epsilon(\lambda_1)$ & $E_r$ 
                & $\lambda_2\!=\!\alpha^2$ & $\epsilon(\lambda_1)$ & $E_r$ 
                \\ 
    \midrule
  1 & 1.11e-3 & 3.4e-3 & 5.73e-3  & 6.76e-4  &  6.18e-2 &  7.16e-3 \\ \hline 
  2 & 1.28e-3 & 6.6e-3 & 5.91e-3 & 4.46e-4  & 2.32e-2  & 3.70e-3 \\ \hline
  3 & 1.54e-3 & 1.24e-2 & 5.17e-3 & 4.85e-4  & 1.70e-2  & 5.31e-3 \\ \hline 
  4 & 1.31e-3 & 1.33e-2 & 5.62e-3 &  5.69e-4  & 7.2e-3  &  5.92e-3   \\ \hline 
  5 & 1.47e-3 & 6.2e-3 & 5.45e-3 &  6.53e-4  &   6.5e-3  &  7.77e-3 \\ \hline 
  6 & 1.32e-3 & 6.1e-3 & 5.09e-3 & 8.41e-4   & 2.04e-2 & 1.04e-2   \\ \hline 
  7 & 6.02e-4 & 1.03e-2 & 7.02e-3 & 8.76e-4   & 5.91e-2 & 1.09e-2   \\ \hline 
  8 & 1.94e-3 & 6.7e-3 & 5.29e-3 & 8.77e-4  & 4.5e-3 & 1.23e-2  \\ \hline 
  9 & 7.32e-4 & 5.6e-3 & 6.36e-3 & 9.17e-4  & 1.51e-2  & 1.76e-2  \\ \hline
  10 & 1.81e-4 & 1.24e-2 & 5.33e-3 & 7.64e-4  & 4.00e-4  & 9.83e-3  \\ \hline 
 Avg
 & 1.31e-3 & 8.30e-3 & 5.70e-3 & 7.11e-3 & 2.12e-2 & 9.09e-3   \\ \hline
$\sqrt{\text{Avg}}$ & 3.62e-2 &  & &                    2.67e-2 &  &   \\ 
     \bottomrule
\end{tabular}

\caption{Ten simulation results  
  with $N_d =$ $ 2000$ training data randomly sampled for each computation.}
\label{rieman-initial-1-table}

\end{table}
}
\begin{figure}[h!]
  \begin{center}
     \includegraphics[width=.7\textwidth]{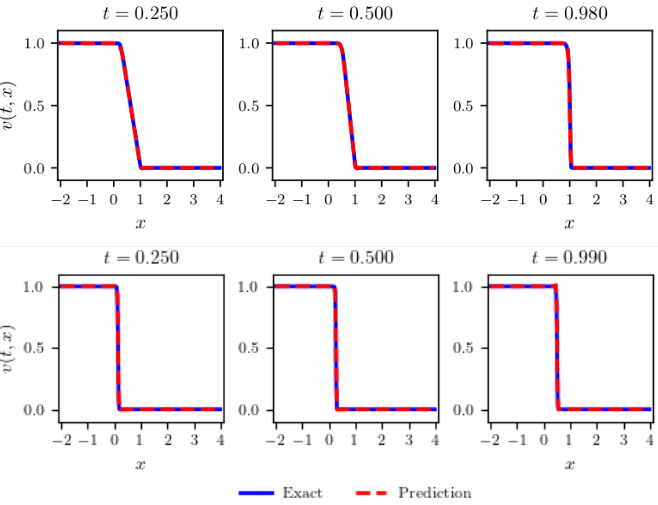} 
  \end{center}
  \caption{\small Top: Evolution with the initial profile (I).
     Bottom: Evolution with the initial profile (II).}
  \label{rieman-initial-1-sol.fig}
\end{figure}

We consider two different initial profiles that develop shocks:
\[
   v_t+vv_x = 0, \ x\in \mathbb{R},\ t\in [0,1) 
\]
with  the initial data, 
\begin{equation}  \label{rieman-initial-1}
\mathrm{(I)}\ v(0,x) = 
    \left\{ \begin{array}{cl}
               1 & \mbox{if}\  x\leq 0 \\
               1-x & \mbox{if}\ 0< x < 1 \\
              0  & \mbox{if}\  x\geq 1  
            \end{array}
            \right. 
\mathrm{and}\,\,  \mathrm{(II)}\  v(0,x) = 
    \left\{ \begin{array}{cl}
               1 & \mbox{if}\  x\leq 0 \\
              0  & \mbox{if}\  x> 0 . 
            \end{array}
            \right.
\end{equation}  
The exact entropy solutions corresponding to the initial data (I) and (II)  
in \eqref{rieman-initial-1} are
\begin{equation} \label{rieman-initial-1-sol}
\mathrm{(I^\prime)}\   v(t,x) = 
    \left\{ \begin{array}{cl}
               1 & \mbox{if}\  x\leq t \\
               \frac{1-x}{1-t} & \mbox{if}\ t< x < 1 \\
              0  & \mbox{if}\  x\geq 1  
            \end{array}
            \right. 
\mathrm{and}\,\, \mathrm{(II^\prime)}\ v(t,x) = 
        \left\{ \begin{array}{cl}
               1 & \mbox{if}\  x\leq \frac{t}{2} \\
              0  & \mbox{if}\  x> \frac{t}{2} , 
            \end{array}
            \right.        
\end{equation}  
respectively. 
The initial profile (I) in \eqref{rieman-initial-1} represents a ramp function with a slope of $-1$, which
creates a wave that travels faster on the left-hand side of $x$ than on the right-hand side.
The faster wave overtakes the slow wave, causing a discontinuity when $t=1$, as we can see
from the exact solution $\mathrm{(I^\prime)}$ in \eqref{rieman-initial-1-sol}.
The second initial data (II) in \eqref{rieman-initial-1} contains a discontinuity at $x=0$. 
Its solution needs a shock fitting just from the beginning. 
Based on the Rankine-Hugoniot condition,
the discontinuity must travel at a speed $x^\prime (t) = \frac{1}{2}$, which we can observe
in the analytical solution $\mathrm{(II^\prime)}$ in \eqref{rieman-initial-1-sol}.
The solution also satisfies 
the entropy condition, which guarantees that it is the unique weak solution for the problem. 
Table \ref{rieman-initial-1-table} shows ten computational results.

  In both cases, the average $\epsilon(\lambda_1)$ is within $3\times 10^{-2}$, 
indicating that the inferred PDE residual reflects the actual Leray-Burgers solutions 
within an acceptable range.
The average value of $\alpha$ with the initial profile (I) was $0.0362$ 
with 
$E_r = 5.7\times 10^{-3}$.
Figure \ref{rieman-initial-1-sol.fig} shows a plot example. 
We can see that the Leray-Burgers solution
captures well the shock wave and maintains the discontinuity at $x=1$ as $t$ evolves to $1$.
Computations with the initial profile (II) resulted in 
$\alpha\approx 0.0267$ on average with
$E_r = 9.1\times 10^{-3}$.
Figure \ref{rieman-initial-1-sol.fig} shows that the Leray-Burgers equation
captures the shock wave as well as its speed $\frac{1}{2}$ per unit time. 
Increasing the training data ($N_d \geq 4000$) did not change the value of $\alpha$ significantly.
      
\subsubsection{Rarefaction Waves}

\begin{figure}[h!]
  \begin{center}
     \includegraphics[width=.7\textwidth]{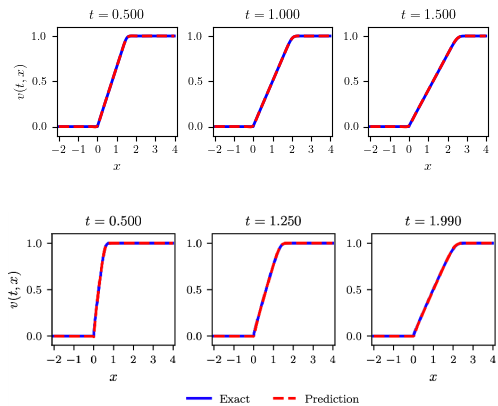} 
  \end{center}
  \caption{\small Top: Evolution with the initial profile (III). 
     Bottom: Evolution with the initial profile (IV).}
  \label{rieman-initial-2-sol.fig}
\end{figure}

We generate a training data set from
the inviscid Burgers equation
\[
 v_t+vv_x = 0, \ x\in \mathbb{R},\ t\in [0,2]
\]
with  the  initial data
\begin{equation} \label{rieman-initial-2}
\mathrm{(III)}\ \ v(0,x) = 
     \left\{ \begin{array}{cl}
               0 & \mbox{if}\  x\leq 0 \\
               x & \mbox{if}\ 0 < x < 1 \\
              1  & \mbox{if}\  x\geq 1  
            \end{array}
            \right. 
\mathrm{and}\ \mathrm{(IV)}\   v(0,x) = 
            \left\{ \begin{array}{cl}
               0 & \mbox{if}\  x\leq 0 \\
              1  & \mbox{if}\  x> 0.  
            \end{array}
            \right. 
\end{equation}  
The rarefaction waves are continuous self-similar solutions, which are
\begin{equation*}
  \mathrm{(III^\prime)}\ \  v(t,x) =    
            \left\{ \begin{array}{cl}
               0 & \mbox{if}\  x\leq 0 \\
               \frac{x}{1+t} & \mbox{if}\  0<x<1+t\\
              1  & \mbox{if}\  x\geq 1+t
            \end{array}
            \right.         
\mathrm{and}\,\, \mathrm{(IV^\prime)}\ v(t,x) = 
             \left\{ \begin{array}{cl}
               0 & \mbox{if}\  x\leq 0 \\
               \frac{x}{t} & \mbox{if}\  0<x<t\\
              1  & \mbox{if}\  x\geq t  
            \end{array}
            \right.  
\end{equation*}  
corresponding to the initial data (III) and (IV)  in \eqref{rieman-initial-2},
respectively.

 In both cases, $\epsilon(\lambda_1)$ is within $10^{-2}$, 
indicating that the inferred PDE residual reflects the Leray-Burgers equations within an acceptable range.
The average value of $\alpha$  are $0.0488$ 
with  $E_r= 1.99\times 10^{-3}$ for the continuous initial profile (III)
and
$\alpha\approx 0.0276$ with 
$E_r = 7.5\times 10^{-3}$ for the discontinuous 
initial profile (IV). 
Figure \ref{rieman-initial-2-sol.fig} shows that the LB equation
captures the rarefaction waves well. 

\subsubsection{Shock and Rarefaction Waves}

We combine the shock and rarefaction profiles:
\begin{equation*} \label{rieman-initial-3}
\mathrm{(V)}\ \ v(0,x) = 
     \left\{ \begin{array}{cc}
               0 & \mbox{if}\  x\leq -1 \\
               1+x & \mbox{if}\ -1< x \leq 0 \\
               1-x & \mbox{if}\ 0< x \leq 1 \\
              0  & \mbox{if}\  x> 1  
            \end{array}
            \right.  
\mathrm{and}\  \mathrm{(VI)}\    v(0,x) = 
            \left\{ \begin{array}{cc}
               0 & \mbox{if}\  x< 0 \\
               1 & \mbox{if}\ 0\leq x \leq 1 \\
              0  & \mbox{if}\  x>1  
            \end{array}
            \right. 
\end{equation*}
\begin{figure}[ht]
  \begin{center}
     \includegraphics[width=.7\textwidth]{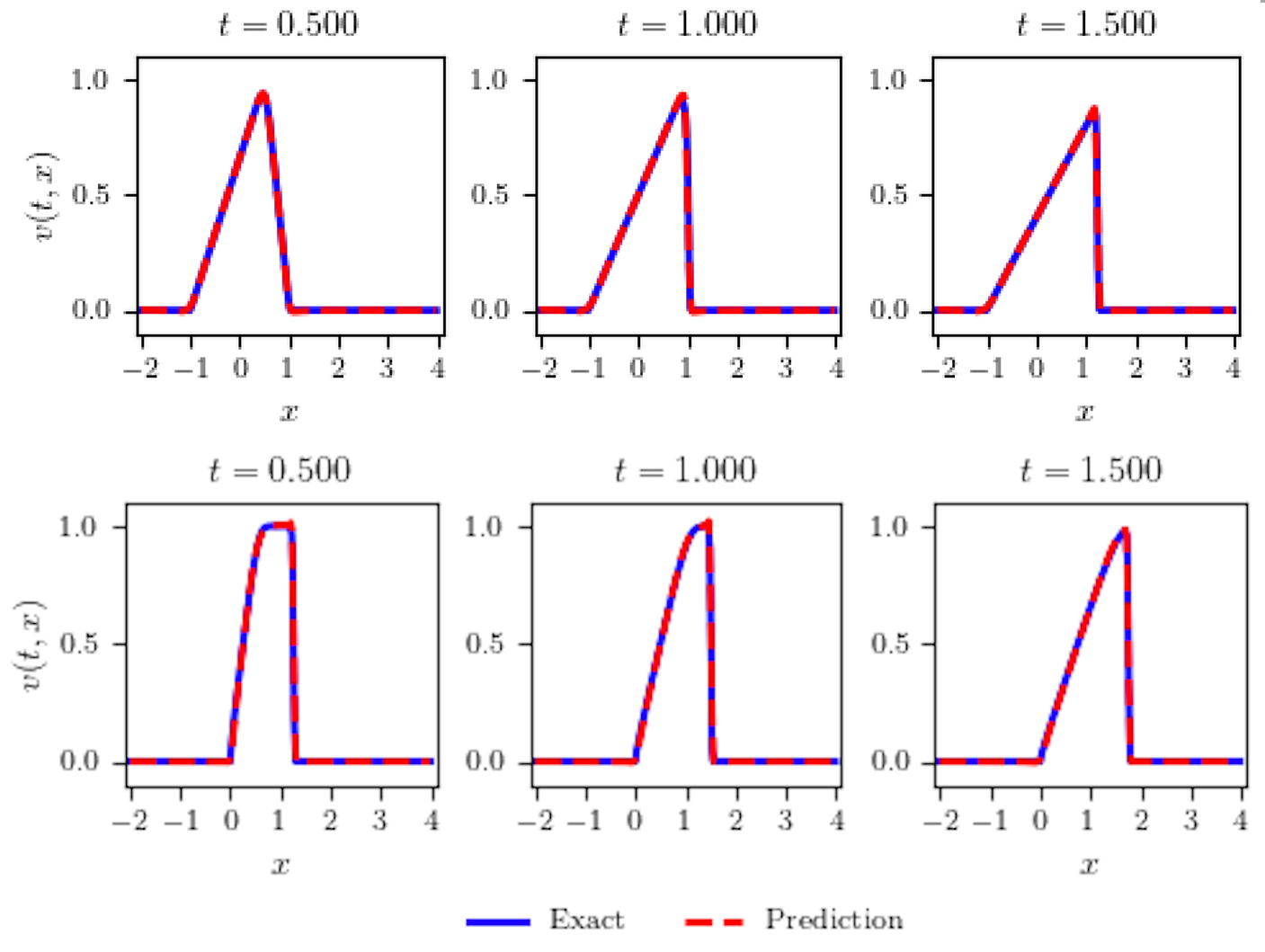} 
  \end{center}
  \caption{\small Top: Evolution with the initial profile (V). 
     Bottom: Evolution with the initial profile (VI).}
  \label{rieman-initial-3-sol.fig}
\end{figure}

In both cases, $\epsilon(\lambda_1)$ is within $10^{-2}$
and the mean values of $\alpha$  are $0.0348$ 
with 
$E_r = 1.95\times 10^{-2}$ for the continuous initial profile (V)
and
$\alpha\approx 0.0316$ with 
$E_r \approx 3.17\times 10^{-2}$ for the discontinuous 
initial profile (VI). 
Figure \ref{rieman-initial-3-sol.fig} shows that the LB equation
captures both shock and rarefaction waves well. 

\subsection{Experiment 2: Viscid Cases}

In this section, we consider the following viscous Burgers equation for a training 
data set:
\begin{equation} \label{viscid1} 
v_t+vv_x = \nu v_{xx}, \ \ \ 
     \forall (t, x)\in (0,T]\times \Omega  
\end{equation}        
 with
\begin{align*}
& \mathrm{(A)}\ \ 
      \left\{\begin{array}{cc}
               \nu =  \frac{0.01}{\pi},  \ T =1,  \Omega = [-1, 1]  &  \\
                v(0,x) = -\sin (\pi x), \  \forall x\in \Omega & \\
                 v(t,-1) = v(t,1) = 0,  \ \forall t\in [0,1] 
                \end{array}
      \right.  \hspace{0.3in} \\[2mm] 
       \mathrm{and}\,\, & \mathrm{(B)}\  \ 
             \left\{\begin{array}{cc}
             \nu = 0.07, \ T\approx 0.4327,\  \Omega = [0, 2\pi] & \\
                v(0,x) = -2\nu\frac{\phi^\prime (x)}{\phi (x)} + 4, \  \forall x\in \Omega & \\
                 \phi(x) = \exp\left(\frac{-x^2}{4\nu}\right) + \exp\left(\frac{-(x-2\pi)^2}{4\nu}\right).
                \end{array}
              \right.   
\end{align*}  
The corresponding LB equation is
\[
v_t + vv_x  = 
           - \alpha^2 v_xv_{xx}.
\]
 
\begin{figure}[ht]
  \begin{center}
    \includegraphics[width=.26\textwidth]{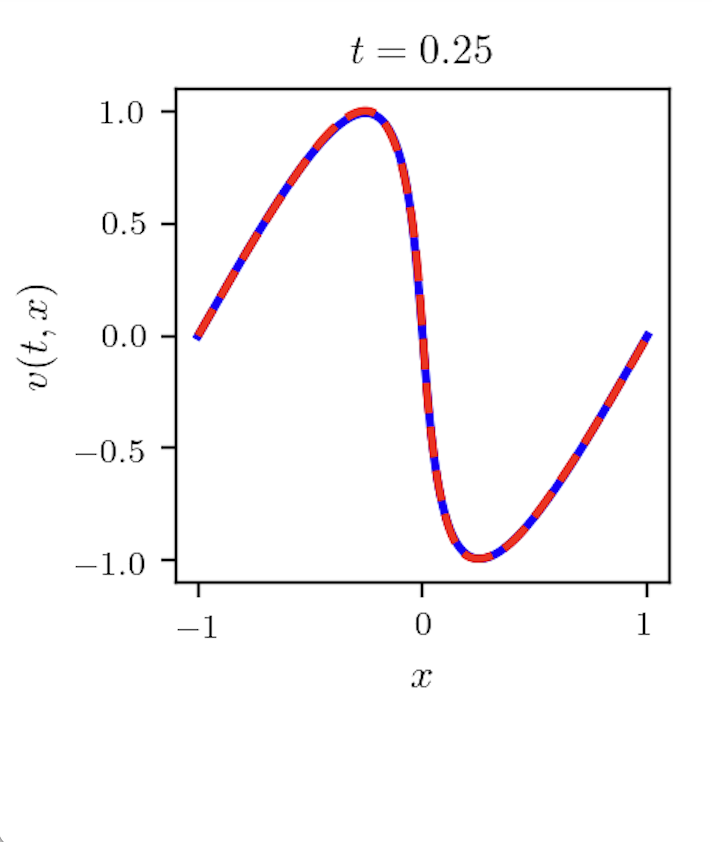}
    \includegraphics[width=.27\textwidth]{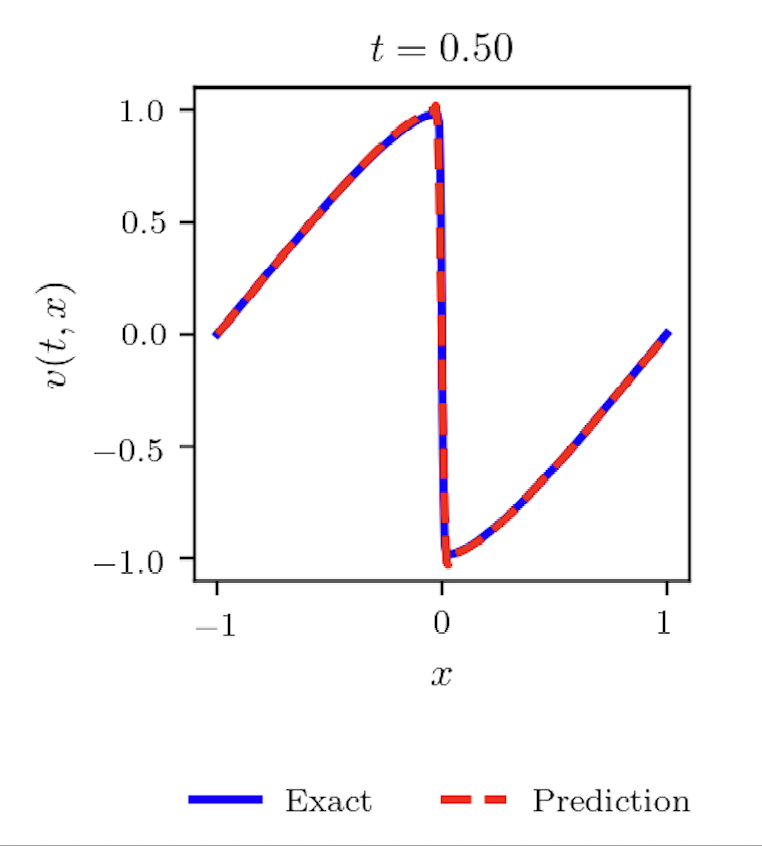}
    \includegraphics[width=.255\textwidth]{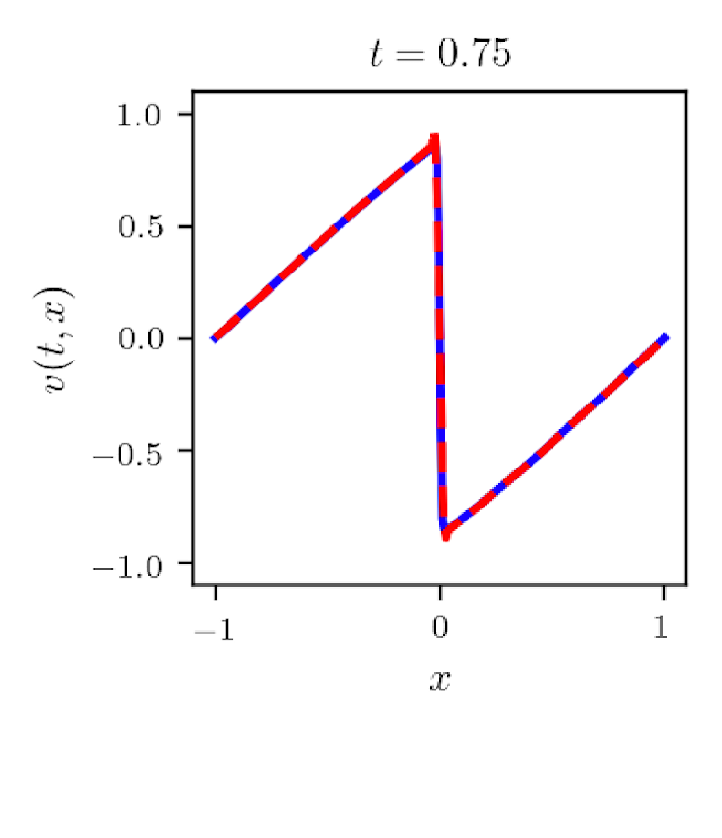}
  \end{center}
  \caption{\small  Example with $N_d = 2000$ for the initial profile (A) }
  \label{viscid1.fig}
\end{figure}

 For the initial and boundary data (A),   
 Rudy et al. \cite{Rudy} proposed the data set that can correctly identify the viscous Burgers equation
 solely from time series data. It contains 101-time snapshots of a solution to the Burgers equation with 
 a Gaussian initial condition propagating into a traveling wave. Each snapshot has 256 uniform spatial grids.
 For our experiment, we adopt the data set prepared by Raissi et al. in 
 \cite{Raissi1, Raissi2} based on \cite{Rudy}, $101\times 256 = 25856$ data points, 
 generated from the exact solution to \eqref{viscid1}.
 For training, $N_d = 2000$ collocation points are randomly sampled 
 and we use the L-BFGS optimizer with a learning rate of 0.8.
 The average of ten experiments is 
 $\alpha = 0.0158$  with $E_r  \approx 3.8\times 10^{-2}$. 
  The computational simulation shows that the equation develops a shock properly 
  (Figure \ref{viscid1.fig}). Note that $\nu = 0.01/\pi 
    \approx 12.7\alpha^2$.

\begin{figure}[h]
  \begin{center}
     \includegraphics[width=.7\textwidth]{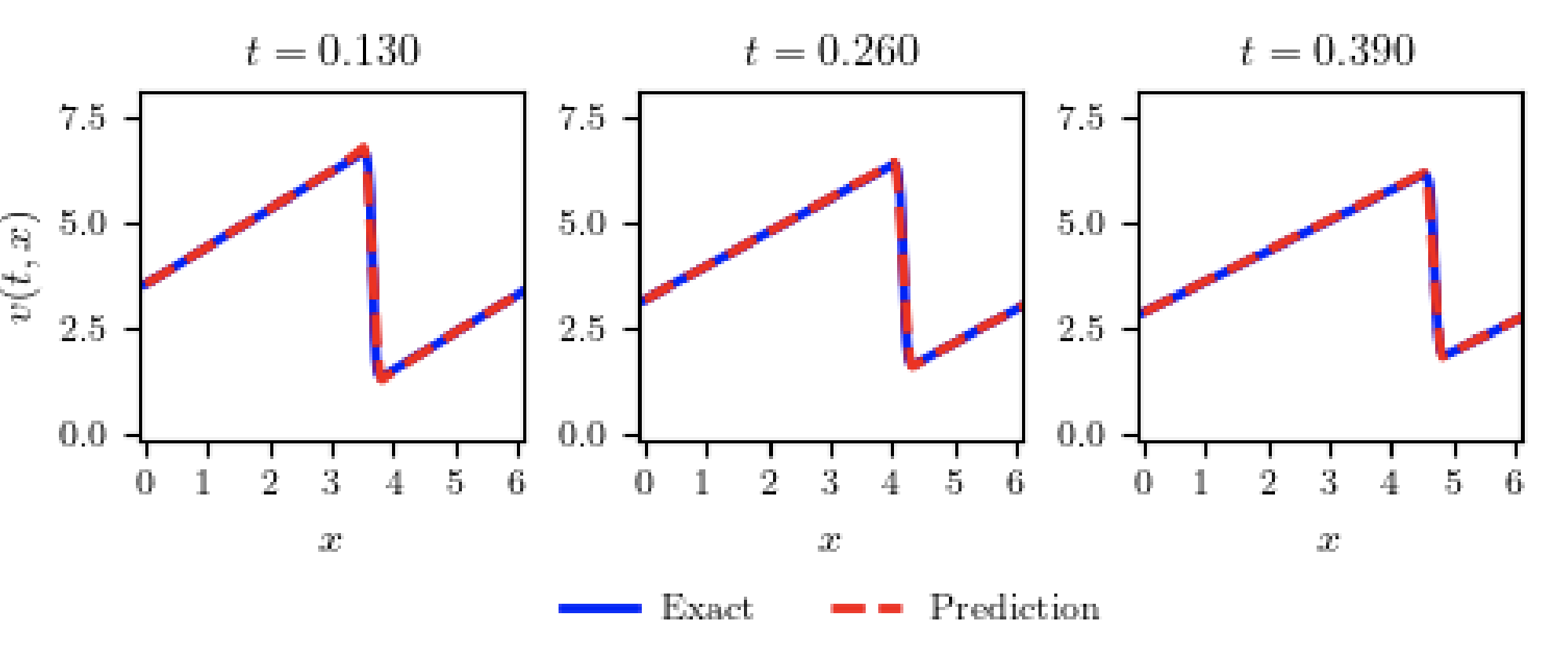} 
  \end{center}
  \caption{\small Example with $N_d = 16000$ for the initial profile (B). }
  \label{EX1-2.fig2}
\end{figure}

For the initial and periodic boundary condition (B), we generate $256\times 500 = 128000$ 
training data 
from the exact solution formula for the whole dynamics over time.
With $N_d = 2000$ training data, the PINN  diverges frequently. 
We experiment with the model with 4000 or more data points to determine an appropriate number of training data.
$L^2$ error remains around $10^{-2}$ for all cases, which does not provide
a clear cut. 
So, we use the absolute error of $\lambda_1$ 
to determine the appropriate number of training data.
For each case of $N_d$, we perform the computation 5 to 10 times (Table \ref{EX1-2.fig1}). 
{\small
\begin{table}[ht]
\centering

\begin{tabular}[t]{ |c|wc{2.4em}|wc{2.4em}|wc{2.4em}|wc{2.4em}|wc{2.4em}|wc{2.4em}|} 
  \toprule
   $N_d$ & 4000 & 6000 & 8000& 10000 & 12000 & 14000   \\ \hline
    $\epsilon(\lambda_1)$ & 0.0207  & 0.0113 & 0.01059  & 0.00965 & 0.00908 & 0.00848 \\ \hline \hline
   $N_d$ & 16000& 18000 & 20000 & 25000 & 30000 &\\ \hline 
    $\epsilon(\lambda_1)$ &  0.0046 & 0.0059
      & 0.00604 & 0.00671 & 0.00611 & \\ 
    \bottomrule
\end{tabular}
\caption{$\epsilon(\lambda_1) =  |1-\lambda_1|$ for various $N_d$ with the initial profile (B).}
\label{EX1-2.fig1}
\end{table}
}
As $N_d$ increases, the results get better and need to do until it reaches the upper limit. 
  Errors between 14000 and 18000 look better than other ranges.
   More than 18000 does not seem to improve the results.
 $N_d = 16000$  (12.5\% of total data) is chosen.
   More than this does not seem to be better. More likely almost the same.
   The average of ten computations is $\alpha\approx 0.0894$ with 
   $E_r = 1.64\times 10^{-2}$.
  Observe that $\nu = 0.07 \approx 8.8\alpha^2$.
Every part of the solution for (B) moves to the right at the same speed, 
which differs from (A) (Fig. \ref{EX1-2.fig2}). In (A), the left side of
a peak moves faster than the right side, developing a steeper middle.
It resulted in a higher value of $\alpha$ with (B) than with (A).  

In summary, we observe that $\nu = 0.01/\pi 
    \approx 12.7\alpha^2$ with the profile (A) and $\nu = 0.07 \approx 8.8\alpha^2$
    with the profile (B).
These results demonstrate that the LB
  equation can capture nonlinear interactions at significantly smaller length scales compared to 
  the viscous Burgers equation. Notably, numerical schemes for the viscous Burgers equation become
  unstable at lower $\nu$ values, whereas the LB equation maintains stability and 
  convergence under these conditions. This observation will be clearer when we 
  compare the forward inferred solutions of two equations 
  in Section  \ref{section-data-driven} (Part C).

\subsection{Experiment 3: The Filtered Vector $u$}\label{section-filtered}

We write Equation \eqref{eq-main-a} in the filtered vector $u_\alpha = u$, which
is a quasilinear evolution equation that consists of the inviscid Burgers equation plus 
$\mathcal{O}(\alpha^2)$ nonlinear terms \cite{BF1, BF2, BF3}:
\begin{equation}\label{eq1.old2}
 u_t + u u_x =   \alpha^2 (u_{txx} +  u u_{xxx}). 
\end{equation}
We compute the equation with the same conditions as in the previous corresponding experiments. 
The results show that the filtered equation \eqref{eq1.old2} also tends to depend on 
the continuity of the initial profile as shown in Table \ref{filtered.table1}.
{\small 
\begin{table}[ht]
\centering

\begin{tabular}[t]{|c|c||c|c|}
  \toprule
    IC & Continuous & IC & Discontinuous \\ \midrule
       I   &  0.0279 & II & {\bf 0.0004} \\ \hline
       III & 0.0469  & IV & 0.0127 \\ \hline
       V   & 0.0469  & VI & 0.0277 \\ 
   \bottomrule
\end{tabular}
\medskip

\caption{Averaged $\alpha$ values for the filtered vector $u$.
  with $N_d = 2000$ and epochs $= 10000$ except the case II. 
 (IC = Initial Condition as in Section 4.) 
 }
\label{filtered.table1}
\end{table}    
 }
When initial profiles contain discontinuities, the $\alpha$ values are much smaller than 
those with continuous initial profiles. 
Compared to the unfiltered equation \eqref{eq-main-a}, 
the $\alpha$ values for the filtered equation \eqref{eq1.old2} 
are smaller, 
which may cause more oscillation in forward inference. 

\begin{figure}[ht]
  \begin{center}
     \includegraphics[width=.7\textwidth]{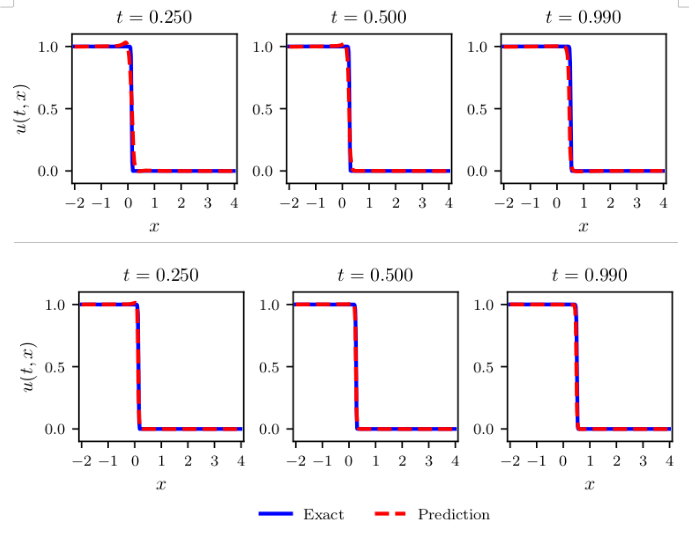}
  \end{center}
  \caption{\small 
  Examples with $N_d = 2000$ for the initial profile (II). 
  Top: epoch=10000. \ Bottom: epoch=50000 }
  \label{EX3-1.fig}
\end{figure}

With the initial profile (II), the parameter $\lambda_1$ for the filtered velocity
is not close to 1 with $\epsilon (\lambda_1) \approx 0.1076$ on average. 
By increasing the number of epochs from 10000 to 50000 we get a better result.
$\lambda_1$ gets closer to 1 with $\epsilon (\lambda_1) \approx 0.0531$, slightly better
relative error and loss, which makes the solution better at later time. 
The oscillation near the discontinuity gets reduced. This verifies that $u$ needs 
very small $\alpha$ values to approximate the inviscid Burgers solution (Figure  \ref{EX3-1.fig}).

In summary, the inverse problem investigations in this section have successfully identified 
a physically meaningful range for the regularization parameter $\alpha$ under various conditions. 
We have established its dependency on the initial data's continuity and also noted the challenges 
of applying this framework to the filtered vector $u$. Having established these foundational 
properties of $\alpha$, we now turn from the inverse problem to the forward inference problem. 
In the following section, we will utilize these findings to develop a data-driven PINN solution 
for the Leray-Burgers equation where $\alpha$ itself is learned dynamically.

\section{Data-Driven Solutions of  the Leray-Burgers Equation}\label{section-data-driven} 

In this section, 
we solve the LB equation  across multiple initial and boundary condition scenarios: 

\[
v_t+vv_x + \alpha^2 v_x v_{xx} = 0, \ x\in \mathbb{R},\ t\in (0,1) 
\]
with
\begin{equation*} \label{rieman-initial-1b}
\mathrm{(I)}\ \ v(0,x) = 
    \left\{ \begin{array}{cl}
               1 & \mbox{if}\  x\leq 0 \\
               1-x & \mbox{if}\ 0< x < 1 \\
              0  & \mbox{if}\  x\geq 1  
            \end{array}
            \right. 
     \mathrm{and} \  \mathrm{(II)}\   v(0,x) = 
    \left\{ \begin{array}{cc}
               1 & \mbox{if}\  x\leq 0 \\
              0  & \mbox{if}\  x> 0 . 
            \end{array}
            \right. 
\end{equation*}
Training utilizes $N_0 = 5000$ initial condition points, $N_b = 5000$ boundary condition points, and $N_r = 20000$ collocation points. These points are adaptively sampled using Latin Hypercube Sampling, with an emphasis on regions exhibiting high PDE residuals or steep solution gradients, particularly in shock regions near discontinuities identified in the initial condition.
Our computational focuses are as follows:
 \begin{enumerate}
    \item {\bf Convergence in $\alpha$}. 
        Whether the PINN solutions converge to those of the inviscid Burgers equation as 
        $\alpha\rightarrow 0^+$.
    \item {\bf Forward inference with  adaptive $\alpha(t)$}.
        Whether the PINN solutions capture the shock and rarefaction waves well and whether the trained $\alpha$ values
           are within the physically valid range.
    \item {\bf Scaling effect of the $\alpha$ parameter relative to the inviscid and viscous Burgers equation}.
 \end{enumerate}

\subsection{The Convergence of the Leray-Burgers Solutions as $\alpha\rightarrow 0^+$}

\begin{figure}[hp]
      \begin{center}
       \includegraphics[width=.9\textwidth]{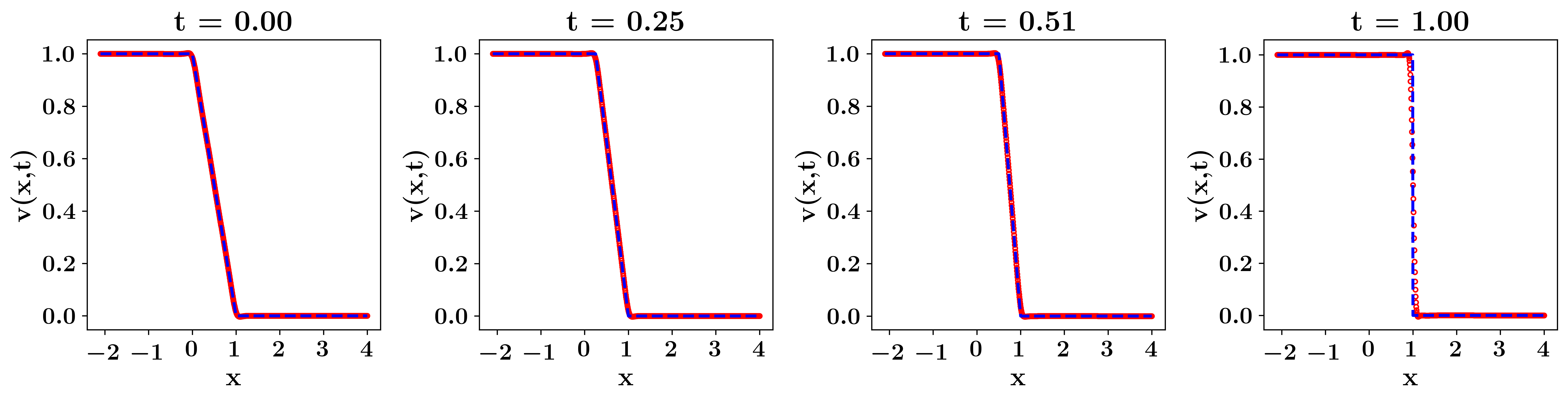}
       \includegraphics[width=.9\textwidth]{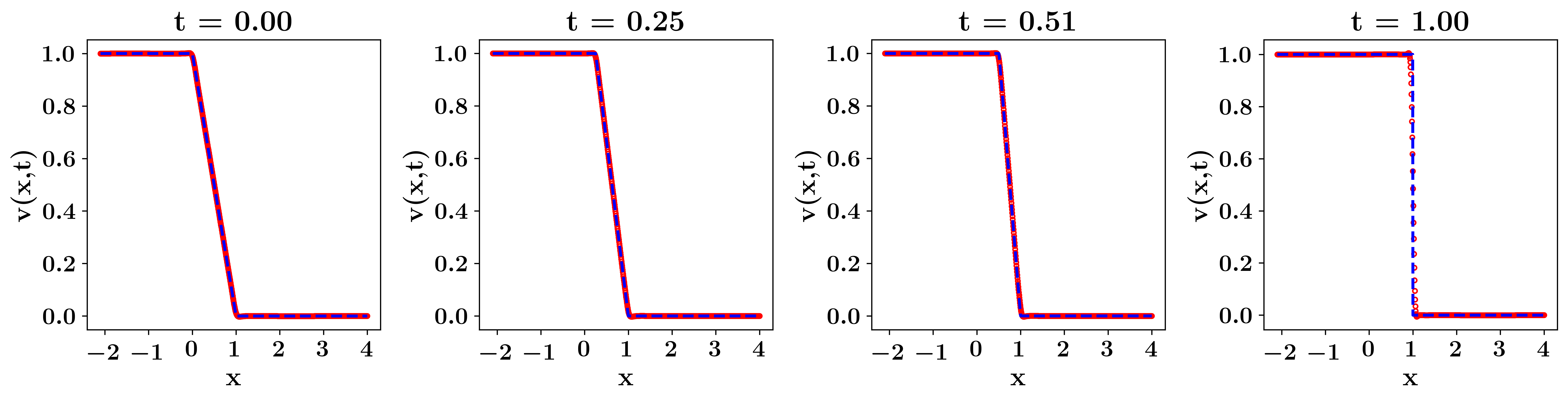}
       \includegraphics[width=.9\textwidth]{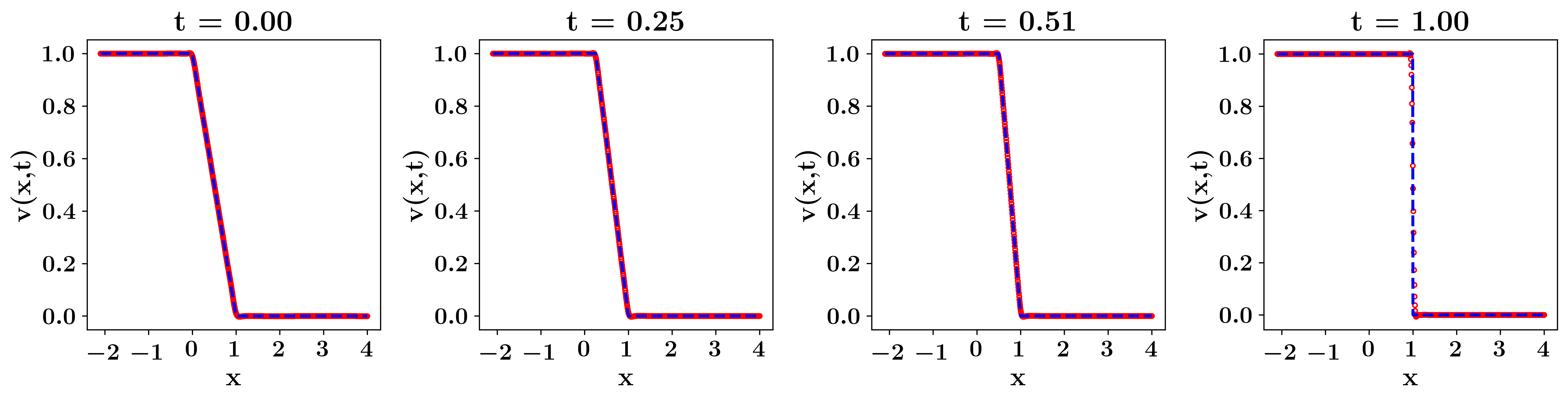}
       \includegraphics[width=.9\textwidth]{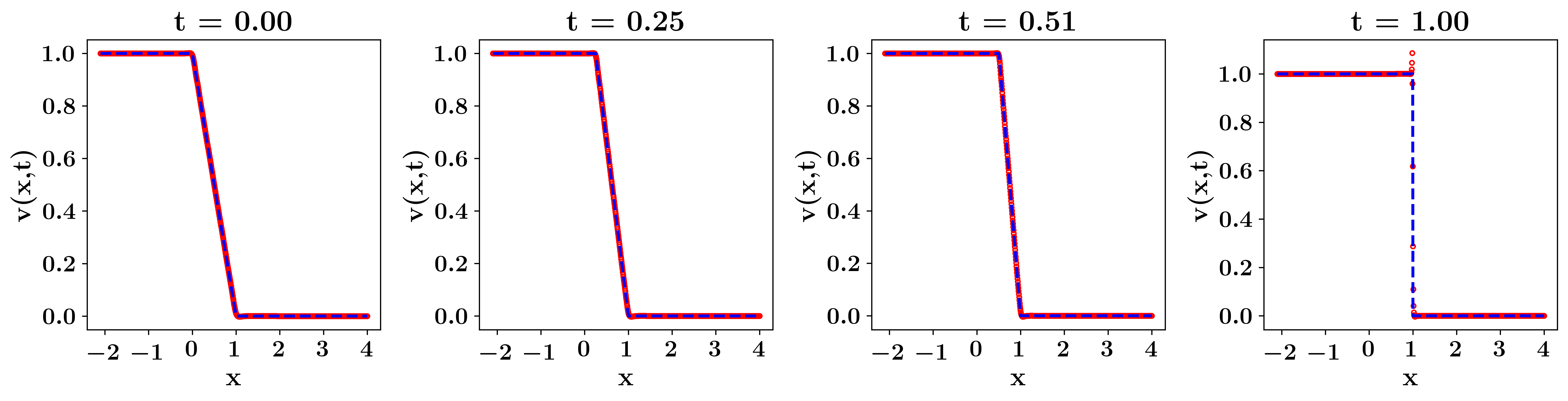}
       \includegraphics[width=.9\textwidth]{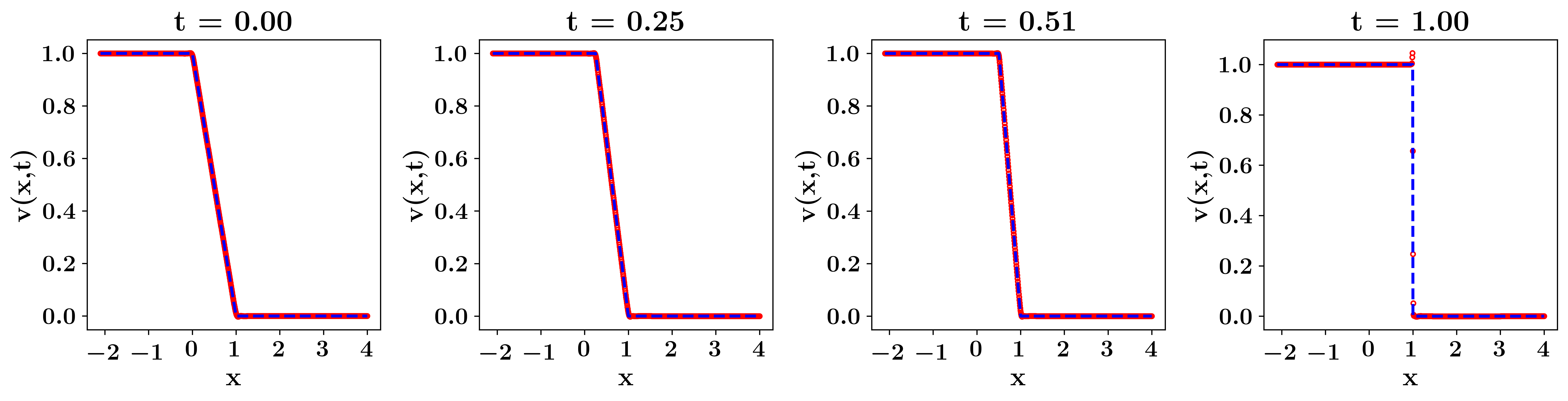}
       
      \caption{\small PINN solution with the initial profile (I) with $N_{0} = 1000$,
        $N_{b} = 1000$,
         and $N_r = 10000$ with epochs = 20000.
         $\alpha = 0.05, 0.04, 0.03, 0.01, 0.0$ from the top
         with  $\alpha = 0$ (inviscid Burgers). The corresponding errors
         $\bar{E}_r = 1.1\times 10^{-2}, 7.9\times 10^{-3}, 5.9\times 10^{-3},
            2.6\times 10^{-3}$, and $3.2\times 10^{-3}$ for $\alpha = 0$,
            respectively.}
      \label{sol1-I.fig}
     \end{center}

\end{figure}

Figure \ref{sol1-I.fig} demonstrates that the Leray-Burgers equation effectively captures 
the shock formation with the continuous initial profile (I) 
within the range of $0< \alpha < 0.05$. 
As $\alpha\rightarrow 0^+$, the LB solution converges to the inviscid Burgers solution (the last graph in Figure \ref{sol1-I.fig}).

 \begin{figure}[h!tbp]
      \begin{center}
      \includegraphics[width=.9\textwidth]{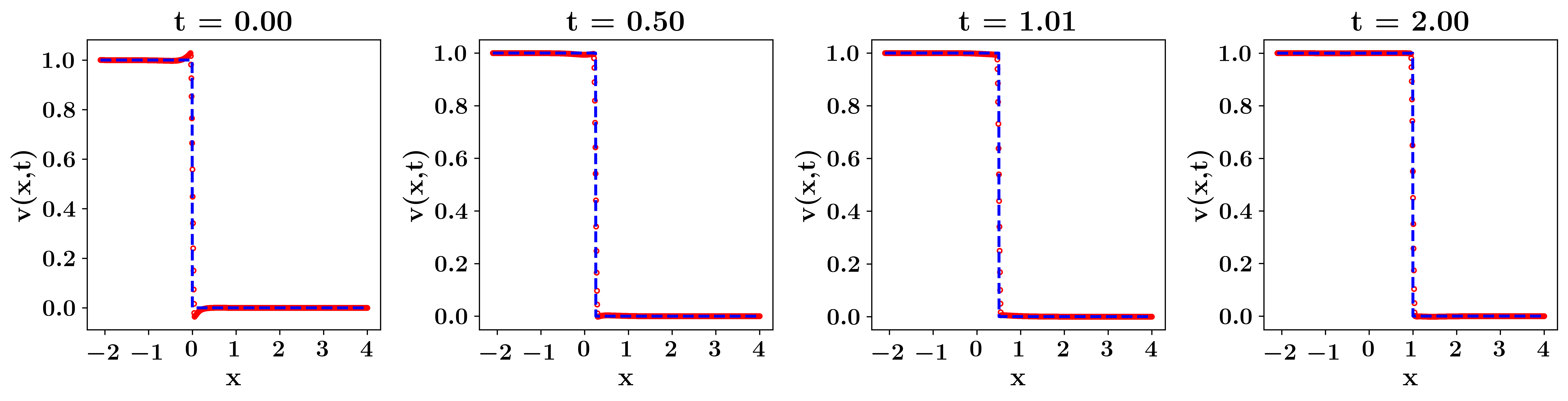}
     \includegraphics[width=.9\textwidth]{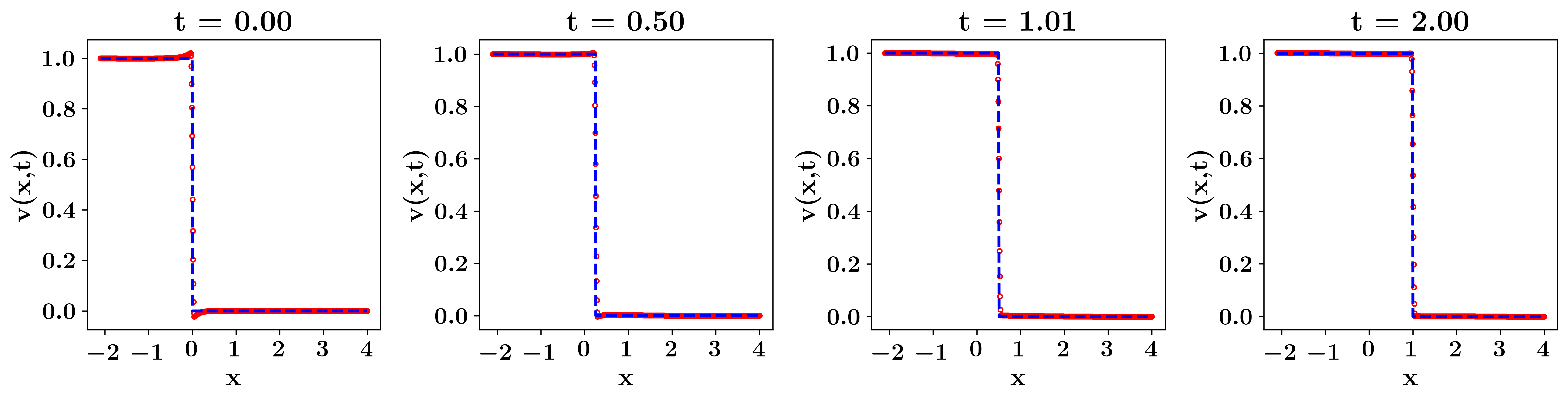}
    \includegraphics[width=.9\textwidth]{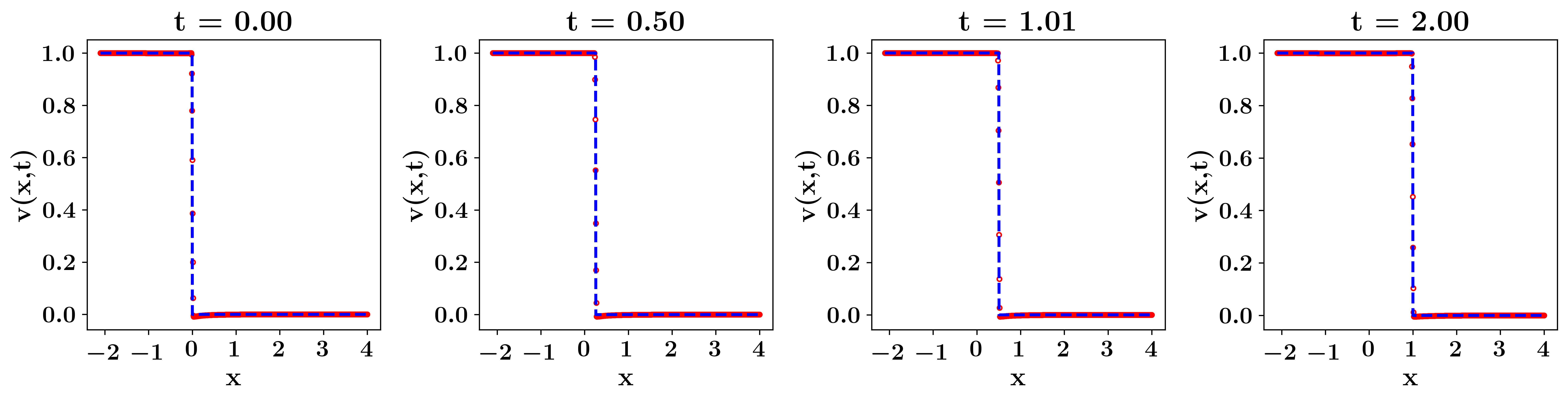}
      \includegraphics[width=.9\textwidth]{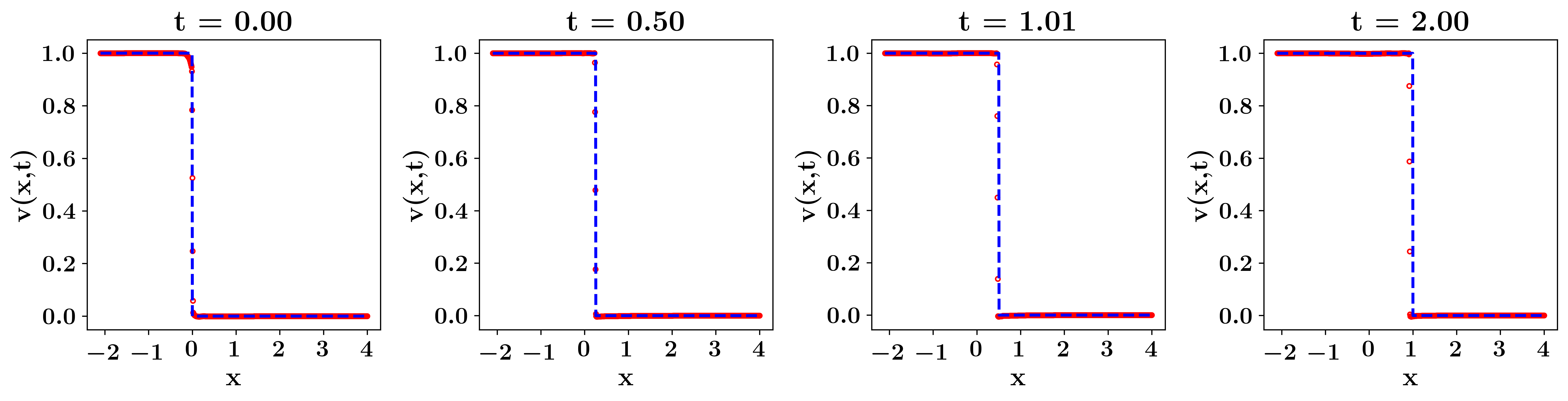}
      \includegraphics[width=.9\textwidth]{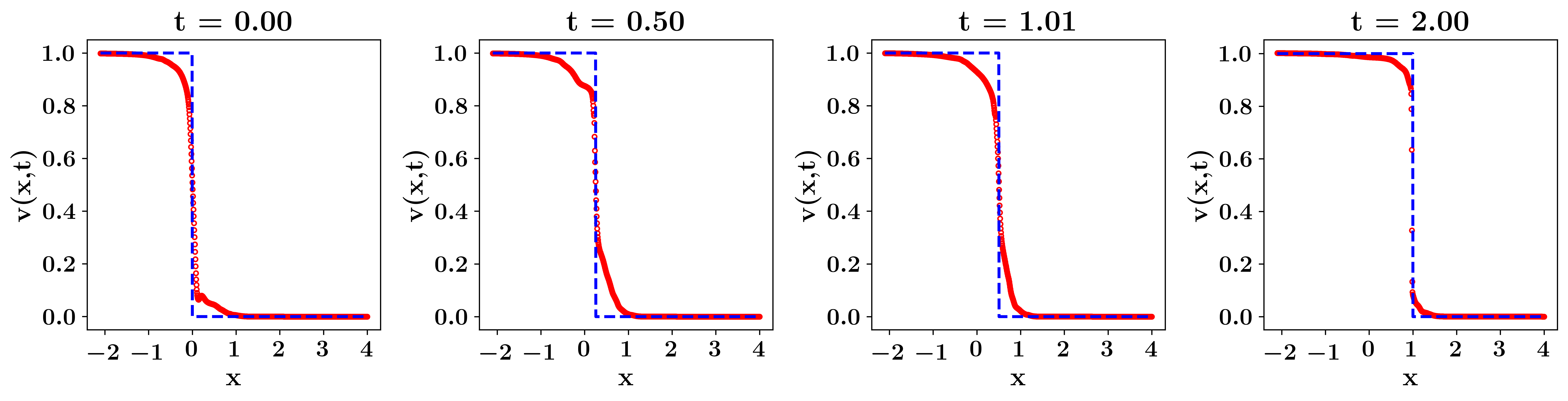}
      \caption{\small PINN solution with the initial profile (II) with $N_{0} = 1000$,
        $N_{b} = 1000$,
         and $N_r = 10000$ with epochs = 20000.
         $\alpha = 0.03, 0.025, 0.015, 0.01, 0.0$ from the top
         with  $\alpha = 0$ (inviscid Burgers). The corresponding errors
         $\bar{E}_r = 4.7\times 10^{-2}, 4.4\times 10^{-2}, 3.4\times 10^{-2},
            9.4\times 10^{-2}$, and $10.0\times 10^{-2}$ for $\alpha = 0$,
            respectively.} 
      \label{sol1-II.fig}
     \end{center}

\end{figure}

 With the discontinuous initial file (II), 
 the Leray-Burgers equation still accurately captures the shock formation 
 within the range of $0.01<\alpha < 0.03$ (Figure \ref{sol1-II.fig}).
 However, the MLP-based PINN generates spurious oscillations near the discontinuity 
 at the beginning. Although the network quickly recovers and fits the oscillations 
 as time progresses, the oscillations worsen, and nonlinear instability arises 
 as the $\alpha$ scale becomes smaller than 0.01, which leads to the deviation
 of the network solution from the actual inviscid Burgers solution 
 (the last graph in Figure \ref{sol1-II.fig}).

\subsection{Forward Inference with Adaptively Optimized $\alpha>0$}

\begin{figure}[h]
      \begin{center}
       \includegraphics[width=.9\textwidth]{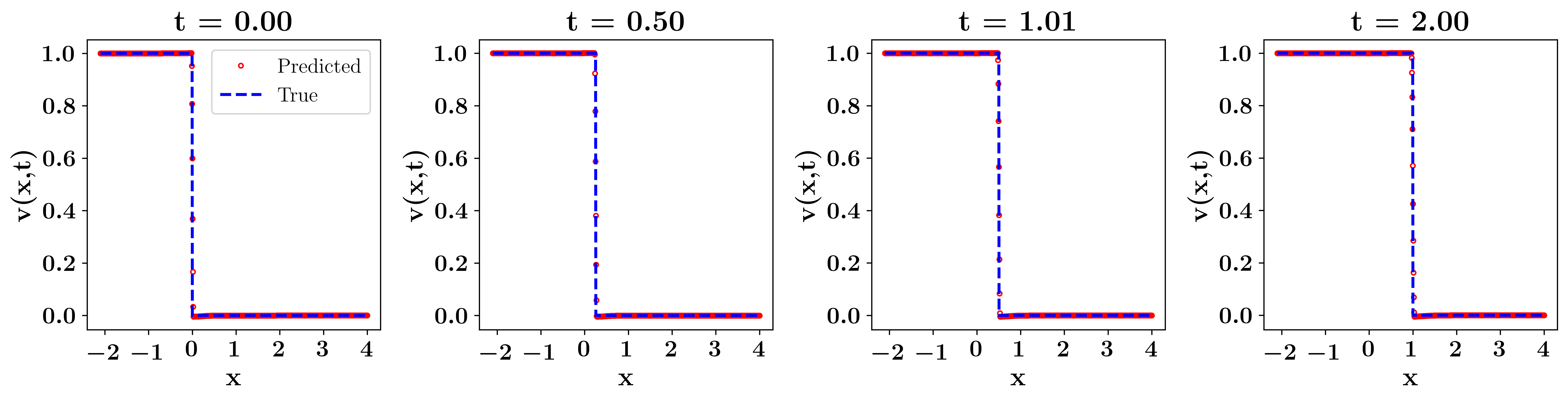}
       \includegraphics[width=.9\textwidth]{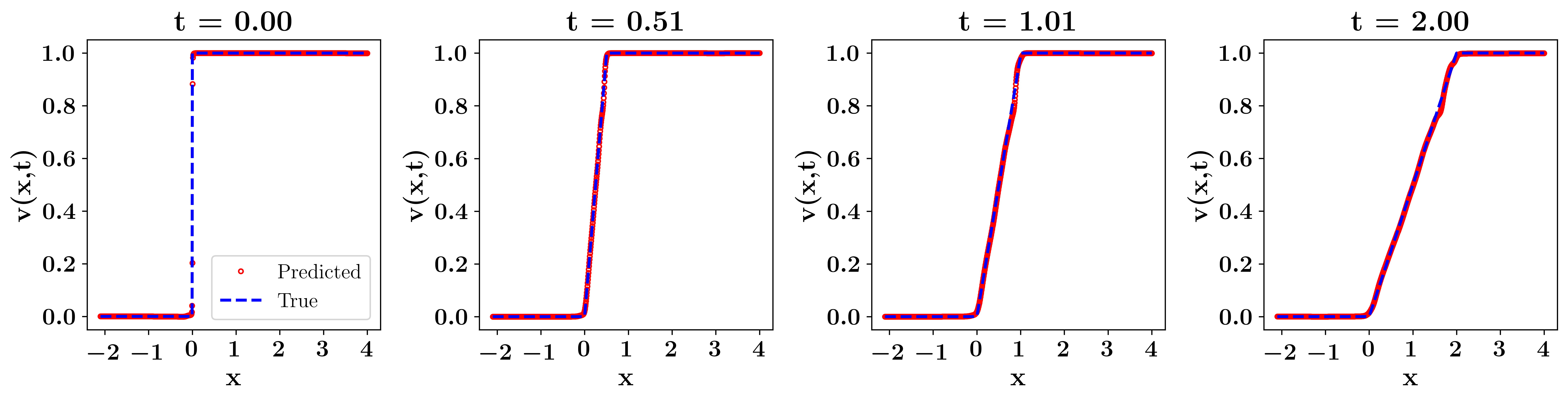}
       \caption{\small Optimal values: (II) $\alpha = 0.0169$ with $\bar{E}_r = 3.5\times 10^{-2}$(top)
           and (IV) $\alpha = 0.0032$ with $\bar{E}_r = 7.9\times 10^{-3}$ (bottom).   }
           \label{optimal-alpha}
     \end{center}
     \end{figure}

In this section, we employ the MLP-based Physics-Informed 
Neural Network (MLP-PINN) to effectively learn the nonlinear operator $\mathcal{N}_\alpha[v]$, 
wherein $v$ represents the primary variable and $\alpha$ denotes a parameter. 
Coutinho et al. \cite{Coutinho} introduced the idea of adaptive artificial viscosity 
that can be learned during the training procedure
 and does not depend on the a priori choice of artificial viscosity coefficient. 
 Instead of incorporating the parameter $\alpha$ in place of the artificial viscosity
 as in \cite{Coutinho}, we set up a dedicated subnetwork, Alpha2Net depicted in Figure \ref{fig1},
 to find the optimal $\alpha(t)$ value. The integration of the subnetwork into the main PINN architecture makes PINN train both $v$ and $\alpha$  to achieve a robust fit with the LB equation. 
 Two examples highlight the ability of the LB equation to capture shock and rarefaction waves 
 as well as the corresponding optimal values of $\alpha$, 
 which are presented in Figure \ref{optimal-alpha}.
     
For the computations, we generated $100\times 1000 = 100000$ training data in 
the domain
$[0,2]\times [-2, 4]$ 
from the corresponding analytical solution for each case.
With $N_0=N_b=1000$, $N_r=10000$, and epochs =20000.
The first graph presents computational snapshots of the system's evolution
with the initial profile (II) over the time interval [0, 2].
The computational outputs are $\mathcal{L}\approx 4.1\times 10^{-4}$ 
and the averaged relative $L^2$-error in time is around $3.5\times 10^{-2}$ with $\alpha\approx 0.0169$. Note that $\alpha$ is the average of trained values of $\alpha(t)$ over time.
The second graph illustrates  snapshots of the evolution of a rarefaction wave
with the initial profile (IV). 
The computational outputs are $\mathcal{L}\approx 1.9\times 10^{-4}$ 
and the averaged relative $L^2$ error in time is around $7.9\times 10^{-3}$ with 
the averaged $\alpha = 0.0032$ in time.

\subsection{The Effect of $\alpha$ Scale in Relation 
to the Inviscid and Viscous Burgers Equations}

When comparing the Leray-Burgers equation \eqref{eq-main-a} with the viscous Burgers equation \eqref{viscid1}, 
the term $\alpha^2 v_x v_{xx}$ in \eqref{eq-main-a} serves as a nonlinear regularization mechanism, 
acting as a substitute for the linear diffusion term in the viscous Burgers equation. 
Unlike linear diffusion,  
    the $\alpha$ term in Equation \eqref{eq-main-a} depends on both the first derivative $v_x$
    and the second derivative $v_{xx}$, 
    suggesting that its smoothing effect is more pronounced in regions with high gradients,
    modulated by the parameter $\alpha$. 
Thus, it is valuable to assess the performance of these two equations 
in relation to the inviscid Burgers equation.
   
Both equations are solved using PINNs with consistent training configurations: 20,000 epochs, fixed weights, and identical network architectures (8 hidden layers, 20 neurons per layer). The key metric for comparison is the \(L^2\) error, which quantifies
the difference between the predicted and exact solutions, with lower values indicating better accuracy.
Computations provide $L^2$ errors for both equations across different values of
\(\alpha\), with \(\nu\) set equal to  \( \alpha^2\) in the viscous Burgers equation
\eqref{viscid1}. The averaged \(L^2\) errors over time are
summarized in Table \ref{tab:results}.
{\small 
\begin{table}[h]
    \centering
    \begin{tabular}{ccccc}
        \toprule
        \(\alpha\) & \(\nu = \alpha^2\) 
                                & LB Average \(L^2\) & VB Average \(L^2\)  \\
        \midrule
        0.025 & 0.000625  & \(2.8025 \times 10^{-2}\) & \( 9.1788\times 10^{-2} \) \\
        0.030 & 0.0009    & \( 2.9933\times 10^{-2}\) &	\(1.2258 \times 10^{-1}\) \\
        0.032 & 0.001024  & \(3.2280 \times 10^{-2}\) & \( 7.8998 \times 10^{-1} \) \\
        0.033 & 0.001089  & \( 3.1334 \times 10^{-2} \) & \( 8.6701\times 10^{-2} \) \\
        0.035 & 0.001225  & \( 3.2484 \times 10^{-2}\) & \( 1.3176\times 10^{-2} \) \\
        \bottomrule
    \end{tabular}
    \caption{Averaged \(L^2\) errors over time for Leray-Burgers (LB)
    and viscous Burgers (VB) equations.}
    \label{tab:results}
\end{table}
} 
For \(\alpha\) values ranging from 0.025 to 0.033 (\(\nu\) from 0.000625 to 0.001089), 
the LB equation consistently outperforms the viscous Burgers equation 
in the averaged \(L^2\) error.
The averaged \(L^2\) error for LB equation remains relatively stable,
ranging from \(2.8025 \times 10^{-2}\)  to \(3.2280 \times 10^{-2}\). 
In contrast, the Burgers equation exhibits higher errors,
ranging from \(7.8998 \times 10^{-2}\) to \(1.2258 \times 10^{-1}\). 
There is no clear monotonic trend, 
indicating variability in the neural network’s ability to approximate the solution.

These results indicate that for small values of $\nu$,  
the viscous Burgers equation is prone to
developing shocks due to its hyperbolic nature.
PINNs may struggle to accurately capture 
these discontinuities. In contrast, the LB equation, through its nonlinear 
regularization effect (dependent on $\alpha$), likely smooths these discontinuities, 
leading to improved accuracy.

\begin{figure}[h!tbp]
      \begin{center}
        \includegraphics[width=.9\textwidth]{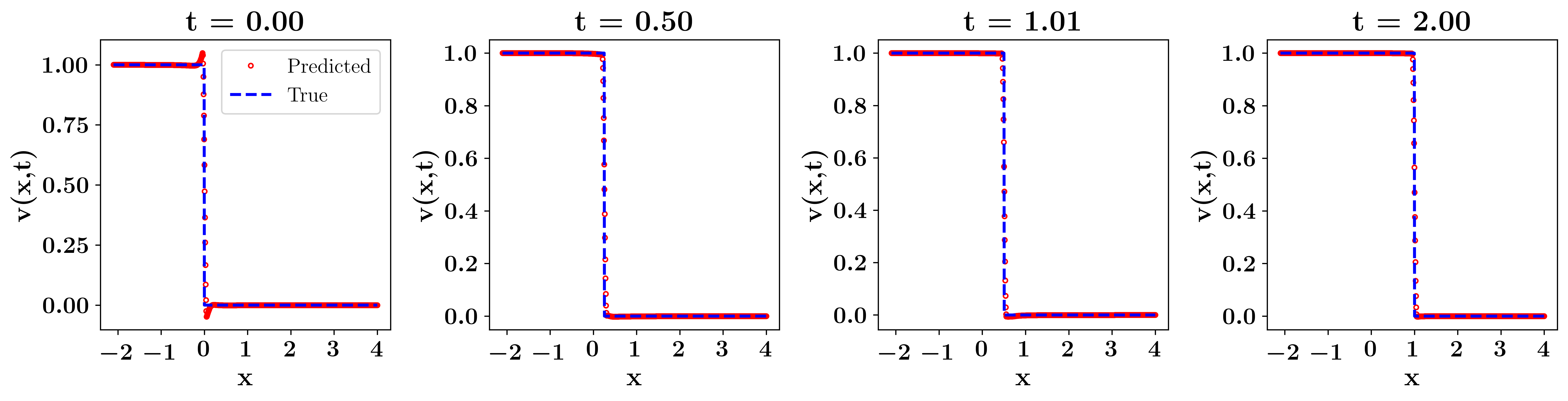}
       \includegraphics[width=.9\textwidth]{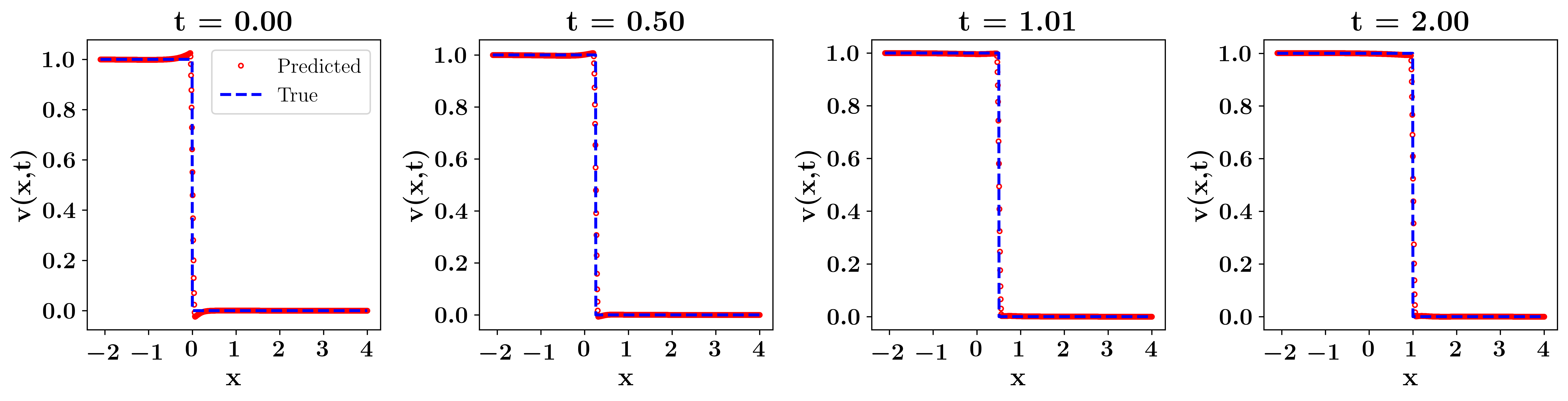}
        \includegraphics[width=.9\textwidth]{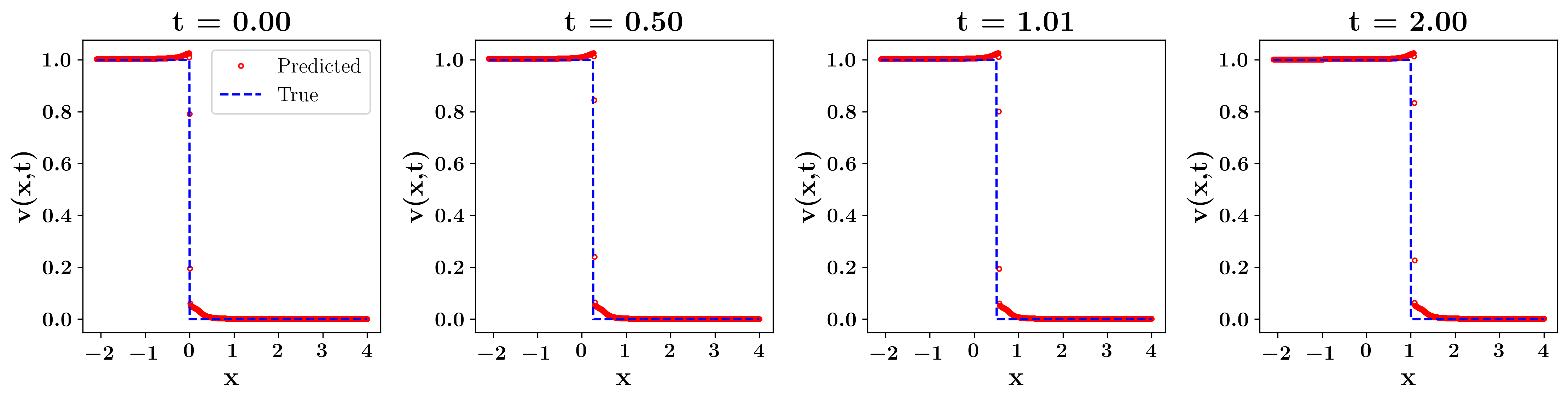}
       \includegraphics[width=.9\textwidth]{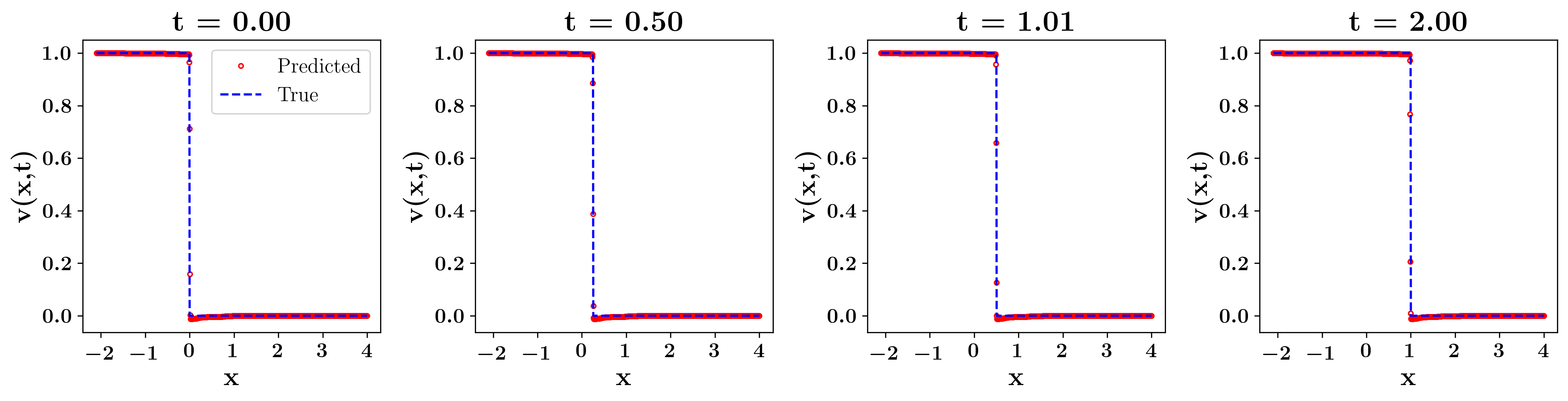}
       \caption{\small Top two rows: 
           Leray-Burgers equation with $\alpha = 0.032$ and $0.035$
           with $\bar{E}_r = 5.1\times 10^{-02}$ and 
           $4.8\times 10^{-02}$, respectively.
           Bottom two rows:
           viscous Burgers equation with $\nu = 0.001024$ and $0.001225$
           with $\bar{E}_r = 1.3\times 10^{-01}$ and 
           $3.9\times 10^{-02}$, respectively.
           }
    \label{Fig9}
     \end{center}
\end{figure}

The data also suggest a tipping point between \(\alpha = 0.032\) 
(\(\nu = 0.001024\)), where the performance of the two models being to shift.
A more definitive transition appears to occur
between \(\alpha = 0.033\) and \(\alpha = 0.035\) (\(\nu = 0.001225\)),
at which point the viscous Burgers equation begins to outperform the LB equation. 
This transition is illustrated in Figure \ref{Fig9}. 

\begin{figure}[ht]
\begin{center}
\includegraphics[width=.45\textwidth]{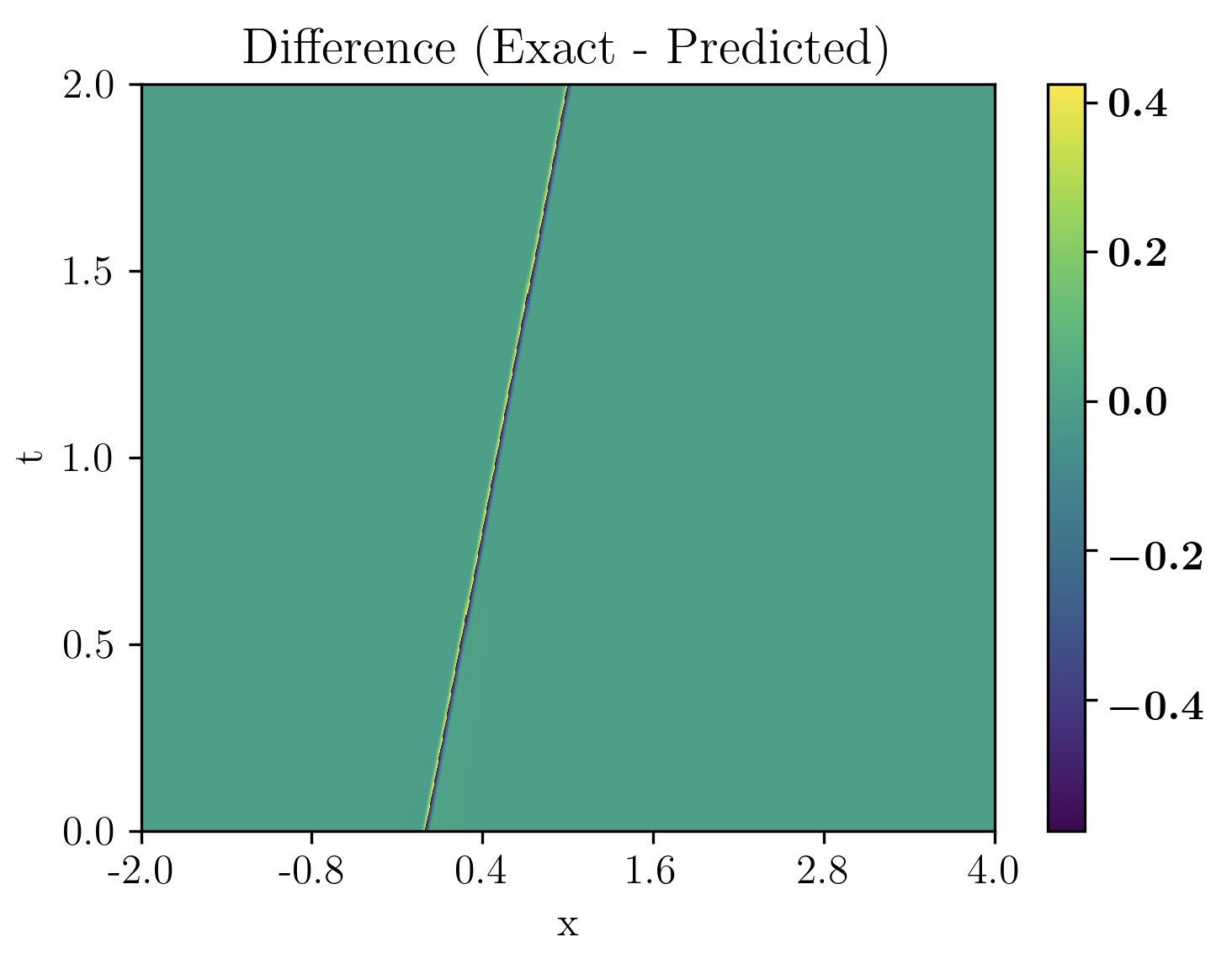}
\includegraphics[width=.45\textwidth]{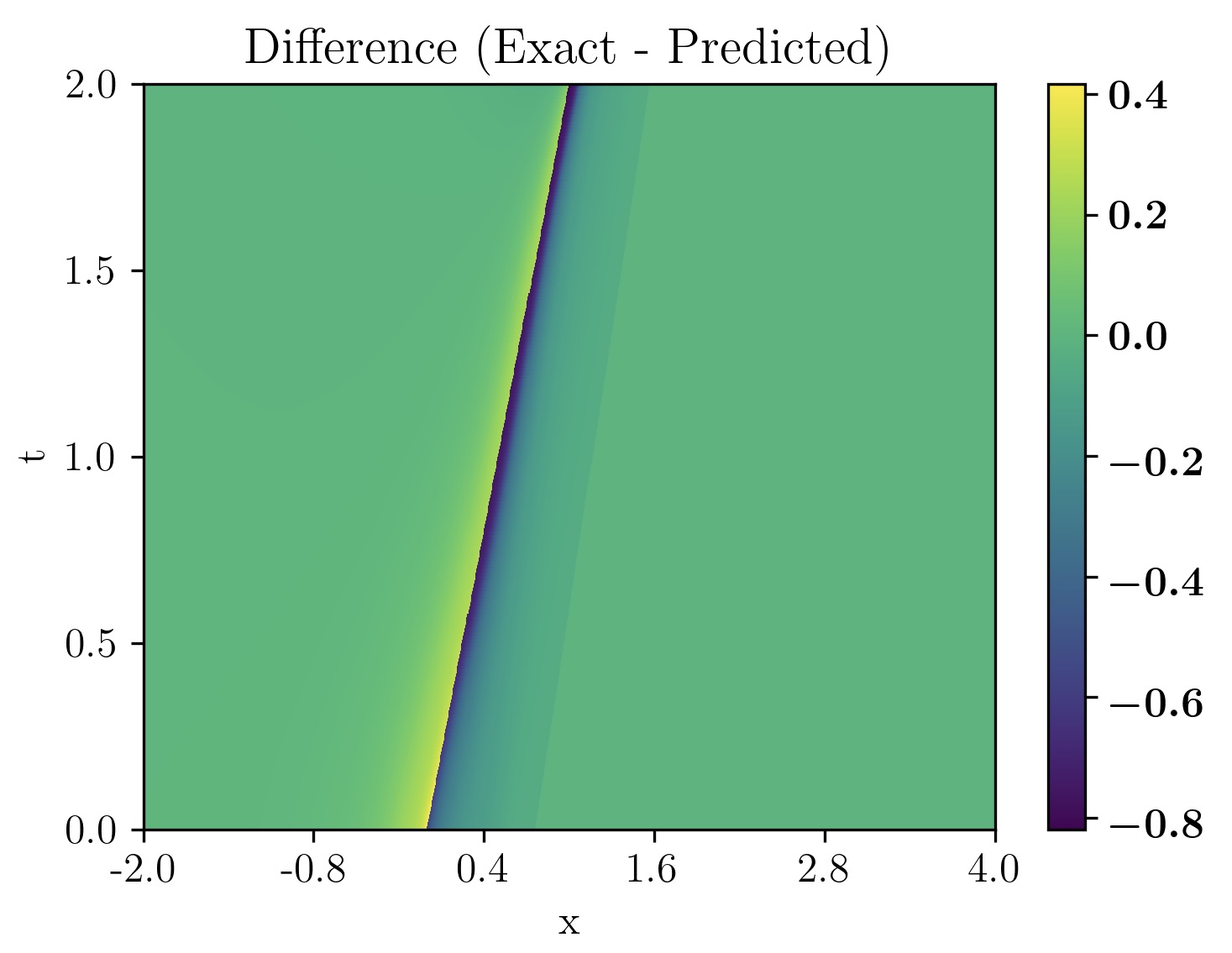}
\caption{\small Heatmaps of the difference between exact and predicted solutions for $\alpha = 0.025$. Top: Leray-Burgers vs. Inviscid Burgers. Bottom: Viscous Burgers vs. Inviscid Burgers.}
\label{Fig10}
\end{center}
\end{figure}

To further examine the differences in solution behavior, Figure \ref{Fig10} presents heatmaps 
of the difference between the solutions of the LB equation and the inviscid Burgers equation, 
as well as the difference between the viscous Burgers equation and the inviscid Burgers 
equation, for $\alpha = 0.025$ ($\nu = 0.000625$). The LB equation exhibits a more gradual 
transition in error distribution, while the viscous Burgers equation shows sharper localized 
discrepancies along the shock region. This suggests that the nonlinear regularization 
in LB equation helps mitigate sharp discontinuities, leading to improved prediction accuracy.

In summary, the parameter $\alpha$ (through $\nu = \alpha^2$) controls the regularization 
strength. Smaller values of $\alpha$ correspond to finer scales where regularization enhances 
accuracy, while larger values increase $\nu$, potentially leading to over-smoothing 
compared to the standard viscous Burgers equation. In practice, the LB equation may be 
preferable for smaller length scales (low $\alpha$), while the viscous Burgers equation may be
more suitable for larger scales (higher $\alpha$). The transition appears to occur 
near $\alpha = 0.035$. These findings emphasize the interplay between physical regularization,
viscosity, and the numerical approximation capabilities of PINNs.

\section{Application to Traffic State Estimation}\label{section-TSE}

This section demonstrates the practical utility of forward inference with the LB equation, 
using the estimated $\alpha$ range to model traffic dynamics efficiently.
Huang et al. \cite{Huang} applied PINNs to tackle the challenge of
data sparsity and sensor noise in traffic state estimation (TSE).
The main goal of TSE is to obtain and provide a reliable description of 
traffic conditions in real time. In Case Study-I in  \cite{Huang},
they prepared the test bed of a 5000-meter road segment for 300 seconds
$\left( (t,x) \in [0, 300]\times [0, 5000]\right)$. The spatial resolution of the dataset is
5 meters and the temporal resolution is 1 second.  The case study was designed to utilize 
the trajectory information data from Connected and Autonomous Vehicles (CAVs) 
as captured by Roadside Units (RSUs), which were deployed every 1000 meters on the road segment
(6 RSUs on the 5000-meter road from $x=0$). 
The communication range of RSU was assumed to be 300 meters, meaning that
vehicle information broadcast by CAVs at $x\in [0, 300]$ can be captured by
the first RSU and the second RSU can log CAV data transmitted at $x\in [700, 1300]$, etc.
More details on data acquisition  and description can be found in \cite{Huang, Huang2}.

In this section, we switch to the differential notation 
$\frac{d\rho}{dt}, \frac{\partial \rho}{\partial x}, \frac{\partial^2 \rho}{\partial x^2}$ 
to avoid confusion with constant parameter notations such as 
$\rho_m$.
   Let $q(t,x)$ denote the flow rate
indicating the number of vehicles that pass a set location in a unit of time and
$\rho(t,x)$ the flow density representing the number of vehicles in a unit road of space.
Then, the  Lighthill-Whitham-Richards (LWR)  traffic model \cite{Huang} is,
for $(t,x)\in \mathbb{R}^+\times\mathbb{R}$,
\begin{equation}\label{LWR}
   \frac{\partial \rho(t,x)}{\partial t}+\frac{\partial q(t,x)}{\partial x} = 0,
\end{equation}
 where $\rho(t,x) = -\frac{\partial N(t,x)}{\partial x}$ and 
    $q(t,x) = \frac{\partial N(t,x)}{\partial t}$. 
    Here $N(t,x)$ is the cumulative flow
    that depicts the number
    of vehicles that have passed location $x$ by time $t$.
Huang, \emph{et all.}, \cite{Huang} adopted the Greenshields fundamental diagram   
to set the relationship between traffic states - density $\rho$, flow $q$, and 
speed $v$:
\begin{equation}\label{Greenshields}
\begin{aligned}
  q(\rho) &= \rho v_f\left( 1 - \frac{\rho}{\rho_m}\right)\\
      v(\rho) &= v_f\left(1 - \frac{\rho}{\rho_m}\right), 
\end{aligned}   
\end{equation}
 where $\rho_m$ is the jam density (maximum density) and $v_f$ is the free-flow speed.
 Substituting the relationship \eqref{Greenshields} into \eqref{LWR} transforms the LWR
 model into the LWR-Greenshield model
 \begin{equation}\label{LWR0}
  v_f\left( 1-\frac{2\rho(t,x)}{\rho_m}\right)\frac{\partial \rho(t,x)}{\partial x}
       + \frac{\partial \rho(t,x)}{\partial t} = 0.
 \end{equation}
 We will just call  it  the LWR model. The equation \eqref{LWR0} is a hyperbolic PDE
 and a second order diffusive term can be added as following,
 to make the PDE become parabolic and 
 secure a strong solution:
 \begin{equation}\label{LWR1}
  v_f\left( 1-\frac{2\rho(t,x)}{\rho_m}\right)\frac{\partial \rho(t,x)}{\partial x}
       + \frac{\partial \rho(t,x)}{\partial t} = \epsilon\frac{\partial^2\rho}{\partial x^2}.
 \end{equation}
 We will call the equation \eqref{LWR1} the LWR-$\epsilon$ model.
  The second-order diffusion term ensures that the solution of PDE is continuous
  and differentiable, avoiding breakdown and discontinuity in the solution.
  Following the same structural idea from \eqref{LWR0} to \eqref{LWR1}
  we add a regularization term to \eqref{LWR0} instead of the diffusion term in
  \eqref{LWR1}:
  \begin{equation}\label{LWR2}
   v_f \!\left( 1\! -\! \frac{2\rho(t,x)}{\rho_m}\right)\frac{\partial \rho(t,x)}{\partial x}
       + \frac{\partial \rho(t,x)}{\partial t} =
        -\alpha^2\frac{\partial\rho}{\partial x}\frac{\partial^2\rho}{\partial x^2}.
  \end{equation}
  We will call Equation \eqref{LWR2}  the LWR-$\alpha$ model.
  We set up the same PINN architecture for computational comparisons of 
  three models, LWR, LWR-$\epsilon$, and LWR-$\alpha$, with
  $v_f = 25\,m/s, \rho_m = 0.15\,vehicles/m$.

\begin{figure}[htbp]
\centering

\begin{tabular}{l c c c}
\toprule
\textbf{Metric} & \textbf{LWR} & \textbf{LWR-$\alpha$} & \textbf{LWR-$\epsilon$} \\
\midrule
Relative Error   & \num{0.107 \pm 0.037} & \num{0.095 \pm 0.016} & \num{0.088 \pm 0.011} \\
Absolute Error   & \num{0.211 \pm 0.114} & \num{0.254 \pm 0.112} & \num{0.153 \pm 0.042} \\
\addlinespace
Training Time (s) & \num[round-precision=1]{333.7 \pm 171.2} & \num[round-precision=1]{158.0 \pm 129.2} & \num[round-precision=1]{336.6 \pm 163.7} \\
\bottomrule
\end{tabular}

\vspace{1.5em} 

\includegraphics[width=\textwidth]{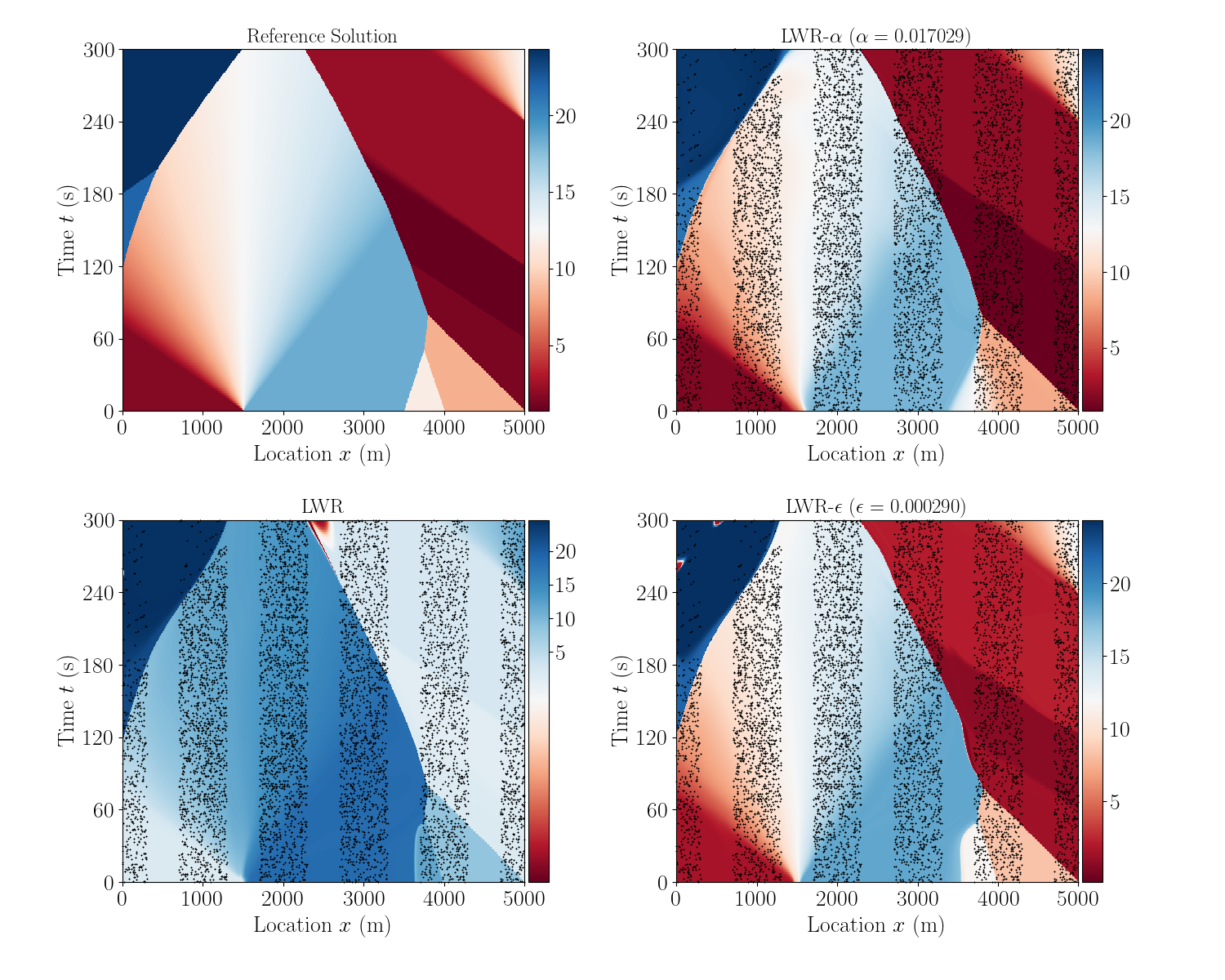}

\caption{
Quantitative and qualitative comparison of models for the Traffic State Estimation (TSE) application. 
\textbf{(Top)} The table summarizes performance metrics (Mean $\pm$ Standard Deviation over 10 runs). 
\textbf{(Bottom)}    \emph{Traffic State Estimation}: 
          The plots visualize the predicted traffic velocity $v(x,t)$ for each model against the reference solution, with black dots indicating the sparse training data locations.
          Shown clockwise from the top left are: the reference speed $v$, 
          the LWR-$\alpha$ estimate ($\alpha=0.017$), 
         the LWR-$\epsilon$ estimate ($\epsilon = 0.00029$), and and the standard LWR estimate.
          Relative errors are approximately $10^{-1}$.} 
\label{fig:tse_summary}
\end{figure}

In our traffic state estimation (TSE) application, 
we demonstrate the practical utility of 
our findings by applying the Leray-type regularization to the LWR model. 
We employed the LWR-$\alpha$ model with a fixed parameter $\alpha = 0.017029$, 
a value chosen to be consistent with the physically meaningful range identified 
in our inverse problem study (Section 4). A new PINN was then trained specifically for this TSE problem.
This parameter value directly addresses our primary objectives of determining
the practical range of $\alpha$ in the Leray-Burgers equation and evaluating the effectiveness
of Physics-Informed Neural Networks (PINNs) in solving the forward inference problem. 
The chosen value of $\alpha$ aligns closely with the estimated range 
of the inverse problem: specifically 0.01 to 0.05 for continuous initial profiles and 
0.01 to 0.03 for discontinuous profiles.

Our empirical calculations demonstrate that all three models—standard LWR, LWR-$\epsilon$, 
and LWR-$\alpha$—provide reasonable approximations of the reference traffic speed $v(t,x)$,
validating their potential for Traffic State Estimation (TSE) as illustrated in 
Figure~\ref{fig:tse_summary}. While the models exhibit broadly comparable accuracy 
in capturing the complex traffic dynamics, the quantitative analysis reveals a critical trade-off 
between accuracy and computational efficiency. The LWR-$\alpha$ model, in particular, 
emerges as the most practical choice. It achieved comparable error rates to the other models 
but in significantly less time. Specifically, when trained on a dataset of 10000 training data 
points and 30000 collocation points for 300000 epochs on a Dell PowerEdge R740 server 
equipped with an NVIDIA Tesla P40 GPU, the LWR-$\alpha$ model required an average of only 158 seconds. 
This is compared to 334 seconds for the standard LWR model and 409 seconds for the LWR-$\epsilon$ model. 
This substantial speed advantage—being more than twice as fast—positions the LWR-$\alpha$ model 
as a highly effective and efficient alternative for TSE applications where timely predictions 
are crucial.

The successful application of the LWR-$\alpha$ model in accurately capturing 
the dynamics of traffic flow 
validates the physical relevance and practicality of our estimated $\alpha$ range. 
Moreover, the robust performance of PINNs in precisely estimating traffic states using this 
model demonstrates their effectiveness in solving the forward inference problem for 
the Leray-Burgers equation. Consequently, our TSE results substantiate both the accuracy of 
our $\alpha$ estimation and the capabilities of PINNs, thereby reinforcing the core findings 
of our study and affirming their potential in real-world applications.

\section{Discussion}

The relationship between the inverse and forward problems is a cornerstone of our approach 
to solving the Leray-Burgers (LB) equation with Physics-Informed Neural Networks (PINNs). 
In the inverse problem, we determine the practical range of the characteristic wavelength parameter 
$\alpha$ that ensures that the LB equation closely approximates the inviscid Burgers solution. 
This range, derived from the training of PINNs on inviscid Burgers data, reflects the values of 
$\alpha$ that maintain the physical fidelity of LB solutions under a variety of initial conditions.

This estimation is not an isolated step,  but directly informs the forward inference process. 
When training PINNs to solve the LB equation, we do not prescribe a fixed 
$\alpha$. Instead, 
$\alpha$ is treated as a trainable parameter, optimized concurrently with the standard PINN parameters
(weights and biases) through a subnetwork called Alpha2Net. To ensure that the optimized
$\alpha$ remains physically meaningful, Alpha2Net enforces a constraint: $\alpha$ must be within the range established by the inverse problem. 
This restriction serves a dual purpose: it prevents the network from converging to nonphysical or suboptimal values of $\alpha$, and it leverages the prior knowledge gained from the inverse problem 
to improve the accuracy and stability of the forward solutions.
For example, if the inverse problem indicates that 
$\alpha$ should range between 0.01 and 0.05 for certain profiles, Alpha2Net ensures that the 
$\alpha$ learned during forward inference adheres to these bounds. 
This linkage guarantees that the solutions to the LB equation not only capture complex phenomena 
like shocks and rarefactions but also remain consistent with the physical constraints established earlier.
Thus, the inverse problem provides an essential scaffold that supports and refines forward inference, 
creating a unified framework for parameter estimation and PDE solution.

While our study demonstrates the utility of PINNs with adaptive $\alpha$ for 
the Leray-Burgers equation, standard MLP-based PINNs face limitations in capturing sharp 
discontinuities such as shocks. As shown in our results (e.g., Figure 9), 
spurious oscillations often arise near discontinuities, 
particularly as the regularization parameter $\alpha$ approaches zero. 
This behavior is often attributed to the spectral bias of MLPs, 
which favor learning low-frequency functions, 
and the difficulty of enforcing pointwise PDE residuals where derivatives are undefined or large.
To address these challenges, the PINN research community has explored various strategies: 
adaptive activation functions and domain decomposition for isolating shock regions 
\cite{Jagtap_Karniadakis};  shock-aware architectures and shock-fitting techniques \cite{Mao};
and weak-form PINNs (e.g., hp-VPINNs, VINO) that bypass strong-form residuals at discontinuities \cite{Kharazmi, Eshaghi}.
A particularly relevant recent development is the Coordinate Transformation PINN (CT-PINN) 
proposed by Chen et al. \cite{Chen_CTPINN_2025}.
The core idea is to transform subdomains, divided along these characteristic curves, 
into regular, simpler domains.
CT-PINNs jointly learn both the characteristic curves and the solution in the transformed domain.
 This approach has shown promise in accurately capturing shock waves 
without significant numerical oscillations by essentially 'straightening out' the characteristics
in the computational domain.

\section{Conclusion}

Computational experiments show that the $\alpha$-values depend on the initial 
data. 
Specifically, the practical range of $\alpha$ spans from 0.01 to 0.05 for continuous initial profiles 
and narrows to 0.01 to 0.03 for discontinuous profiles.   
We also note that
the Leray-Burgers equation  in terms of the filtered vector $u$ does 
not  produce
    reliable estimates of  $\alpha$.  
    When approximating the filtered solution $u$ with commendable precision, 
    MLP-PINN necessitates a more extensive dataset, and the range of 
    $\alpha$ values  for $u$  
    appears confined, between 0.0001 and 0.005. 
  Nonetheless, the MLP-PINN's attempts with $u$ encounter challenges in converging to the true Burgers solutions.
   Thus, it is evident that the equation formulated in the unfiltered vector field $v$ offers a better approximation 
   to the exact Burgers equation.    
    
  In practical terms, treating $\alpha$ as an unknown variable becomes a prudent strategy. 
  By endowment $\alpha$ with learnable attributes along with network parameters, MLP-PINNs can be structured to reveal $\alpha$ within a valid range during the training process,
potentially improving accuracy.   Nevertheless, the MLP-PINN does generate spurious oscillations near discontinuities inherent 
   in shock-inducing initial profiles. 
   This phenomenon thwarts the PINN solution from aligning with an exact inviscid Burgers solution as $\alpha\rightarrow 0^+$.

    Our application to Traffic State Estimation (TSE) also validates the practical utility of 
    the LWR-$\alpha$ model, which is based on the Leray-Burgers equation. 
    These experiments highlight that capturing the nonlinear characteristics of traffic flow is 
    critical for accurate state estimation, a task for which the Leray-type regularization is 
    well-suited. The LWR-$\alpha$ model emerges as the more practical choice by offering 
    a superior balance of performance and efficiency. It surpasses the diffusion-based LWR-$\epsilon$
    model in computational speed—being more than twice as fast in our tests—while its inherent 
    nonlinear regularization effectively captures the complex flow dynamics. 
    This combination of robust physical alignment and significant computational efficiency makes 
    the LWR-$\alpha$ model a compelling and viable alternative for real-time TSE applications.





\section*{Statements and Declarations}

 \noindent {\bf Competing Interests}: On behalf of all authors,
  the corresponding author states that 
there is no conflict of interest.
\smallskip

\noindent {\bf Data Availability}: 
The code and data we used to train
and evaluate our models are available at 
\begin{center}
\href{https://github.com/bkimo/PINN-LB}{https://github.com/bkimo/PINN-LB}.     
\end{center}
The data for traffic state estimations
generated by Huang et al. \cite{Huang, Huang2} 
is available at 
\begin{center}
\href{https://github.com/arjhuang/pise}{https://github.com/arjhuang/pise}.
    
\end{center}

\section*{Acknowledgment}
The second author was supported by the Research Grant of Kwangwoon University in 2022 and by the National Research Foundation of Korea(NRF) grant funded by the Korea government(MSIT) (No. 2021R1F1A1058696).
The third author gratefully acknowledge the Advanced Technology and Artificial Intelligence Center at American University of Ras Al Khaimah for providing their GPU computing resources to support this research.


\bibliographystyle{plain} 
\bibliography{KCL-bibliography}


   \end{document}